\documentclass[useAMS,usenatbib]{mnras}
\usepackage{graphicx}
\usepackage{longtable}
\usepackage{graphicx}
\usepackage{hyperref}
\usepackage{amsmath}
\title[IFU spectroscopy of southern PNe: III]
  {IFU spectroscopy of Southern Planetary Nebulae: III}
\author[A. Ali et al.]
  {A. Ali,$^{1,2}$ M.A. Dopita,$^{1,3}$ H.M. Basurah,$^1$ M.A. Amer,$^{1,2}$ R. Alsulami,$^1$ \& A. Alruhaili,$^1$  \\
  $^1$Astronomy Dept, Faculty of Science, King Abdulaziz University, 21589 Jeddah, Saudi Arabia \\
  $^2$Department of Astronomy, Faculty of Science, Cairo University, 12613 Giza, Egypt \\
  $^3$Research School of Astronomy and Astrophysics, Australian National University, Cotter Rd., Weston ACT 2611, Australia
  }
\date{Released 2016 Xxxxx XX}

\pagerange{\pageref{firstpage}--\pageref{lastpage}} \pubyear{2016}

\def\LaTeX{L\kern-.36em\raise.3ex\hbox{a}\kern-.15em
    T\kern-.1667em\lower.7ex\hbox{E}\kern-.125emX}

\begin{document}

\label{firstpage}

\maketitle

\begin{abstract}
 In this paper we describe integral field spectroscopic observations of four southern Galactic Planetary Nebulae
 (PNe), M3-4, M3-6, Hen2-29 and Hen2-37 covering the spectral range; 3400-7000{\AA}.
 We derive the ionisation structure, the physical conditions, the chemical compositions and the kinematical
 characteristics of these PNe and find good agreement with previous studies that relied upon the long-slit technique in their co-spatial area.
 From their chemical compositions as well as their spatial and kinematic characteristics, we determined that  Hen2-29 is of
 the Peimbert Type I (He and N rich), while the other three are of Type II. The strength of the nebular He II line
 reveals that M3-3, Hen2-29 and Hen2-37 are of mid to high excitation classes while M3-6 is a low excitation planetary nebula (PN).
 A series of emission-line maps extracted from the data cubes were constructed for each PN to describe
 its overall structure.  These show remarkable morphological diversity. Spatially resolved spectroscopy of
 M3-6, shows that the recombination lines of C\,II, C\,III, C\,IV and N\,III are of nebular origin, rather
 than arising from the central star as had been previously proposed. This result increases doubts regarding
 the weak emission-line star (WELS) classification raised by \citet{Basurah16}. In addition, they reinforce
 the probability that most genuine cases of WELS are arise from irradiation effects in close binary central stars \citep{Miszalski09}.
\end{abstract}


\section{Introduction}
The vast majority of the spectroscopic studies of PNe up to the
present day have relied upon long slit spectroscopic techniques.
However these measurements sample only a portion of the complete
nebula, and are of necessity weighted towards the high-ionisation
regions around the central star (CS). An accurate determination of
the physical and chemical nebular parameters, as well as the
determination of their global parameters requires a knowledge of
their integrated spectra and spatial structure.  The advent of
Integral Field Units (IFUs) now provides an opportunity to obtain
this data, and to build fully self-consistent photoionization models
for PNe. The IFU technique as applied to PNe was pioneered by
\citet{Monreal-Ibero05} and \citet{Tsamis07}. Recently, detailed
physical and morpho-kinematical studies using optical IFU data have
been taken by \citet{Danehkar15} and \citet{Danehkar15x} to study
the planetary nebulae Hen 3-1333, Hen 2-113 and Th 2-A.

The Wide Field Spectrograph (WiFeS) instrument mounted on the 2.3m
ANU telescope at Siding Spring Observatory \citep{Dopita07,Dopita10}
offers the ability to perform such IFU spectroscopy, since it is
capable of reaching a spatial resolution of 1.0\arcsec, a spectral
coverage of 3200--8950\AA\ and a spectral resolution of $R=7000$.
Combined with a field of view of 25\arcsec x 38\arcsec, it is very
well-suited to integral field spectroscopy of compact PNe. The first
paper in this series by \citet{Ali15b} used WiFeS to study the
large, evolved and interacting planetary nebula PNG 342.0-01.7,
generating an IFU mosaic to cover the full spatial extent of the
object. In the second, \citet{Basurah16} provided a detailed
analysis of four highly excited non-type I PNe which casts doubt on
the general applicability of the WELS classification. Here, and in
upcoming studies, we aim to further exploit the capabilities of the
WiFeS instrument to provide a new database on previously studied,
bright and compact PNe. Specifically our objectives are to:
\begin{itemize}
\item{Create emission-line maps for PN in any diagnostic emission line within its spectral range.}
\item{Provide integrated spectra of the whole PN, and if possible, of its exciting star.}
\item{Analyse these spectra both in the forbidden and recombination lines to derive chemical
abundances and to understand any differences between results obtained here and through long-slit observations of the same objects.}
\item{Determine expansion velocities and the kinematic nature of the PN.}
\item{Build self-consistent photoionisation models to derive abundances, physical conditions
within the nebula, to determine distances, and to place the PNe on the Hertzsprung-Russell
Diagram in order to derive central star masses, the evolutionary status and the nebular age.}
\end{itemize}

Emission-line mapping of PNe has previously been obtained through
the technique of narrow-band interference filter imaging, e.g.
\citet{Manchado96, Gorny99, Hajian07, Miranda10}, \citet{Aller15}
and \citet{Garcia16}. The difficulty here is that each image is very
expensive in observing time, and only one emission line can be be
imaged at once. On the other hand, IFU spectroscopy makes available
the narrow band images of all lines which are observed with
sufficient signal to noise ratio. Emission line maps and line ratio
maps allow us to structural details and physical processes. For
example,[O\,III]/H$\alpha$ ratio maps are useful to study the
variation of ionization and chemical abundances and also to look for
signatures of collimated outflows and shocks \citep{Guerrero13},
[SII]/H$\alpha$ ratio maps are particularly sensitive to shocked
regions \citep{Akras16a,Akras16b}, whereas HeII/H$\alpha$ ratio maps
identify the very high ionization regions \citep{Vazquez12}.

The main objective of the current paper is to study four Galactic
planetary nebulae -- M3-4, M3-6, He2-29 and Hen2-37 -- which have so
far received relatively little attention. In this paper we will use
WiFeS data to examine the correlations between nebular morphology
and excitation. Furthermore, the WiFeS instrument is ideally suited
to study spatial variations in nebular parameters, such as
extinction, electron temperature, density and ionic abundances.

\citet{Gorny99} imaged Hen2-29 and Hen2-37 in a narrow band
H$\alpha$ filter and determined angular sizes of 20\arcsec
$\times$16\arcsec and 27\arcsec $\times$ 24\arcsec, respectively.
\citet{Corradi98} classified Hen2-29 \& Hen2-37  nebulae as
elliptical PNe with modest ellipticity (major to minor axis length
ratio $\leq$ 1.3) and outer irregular contours. Searching the
literature and the ESO archive, it would appear that no narrow band
images are available for M3-4 and M3-6. However, spectroscopic
studies of all four PNe are to be found distributed across several
papers, e.g., \citet{Milingo02}, \citet{Martins00},
\citet{Maciel99}, \citet{Kingsburgh94}, \citet{Chiappini94} and
\citet{Perinotto04}.

In this paper, we present excitation maps, integral field
spectroscopy and an abundance analysis of these four PNe. The
observations and data reduction are described in Section 2, while
the emission-line maps are presented and discussed in Section 3. In
Section 4, we use the spectrophotometry to derive the physical
conditions, ionic and elemental abundances, and excitation class
determinations. Section 5, gives kinematical signatures such as the
expansion and radial velocities and the distances. In Section 6, we
provide a discussion of the classification of the CS in M3-6, and
our conclusions are given in Section 7.
\begin{table*}
 \centering
 \small
   \caption{The observing log}
    \label{Table1}
   \scalebox{0.95}{
  \begin{tabular}{llcccccl}
 \hline
   Nebula name  & PNG number & No. of  & PA ($\deg$) & Exposure  & Date   & Airmass & Standard Star\\
    & & frames & & time (s) &  &   & \\
   \hline \hline
 M3-4 & PN G241.0+02.3  &  3 & 90 &600 & 31/3/2013 & 1.01 & HD 111980  \& HD 031128 \\
 M3-6 & PN G253.9+05.7    &  6 & 0 & 300 & 31/3/2013 &  1.01 & HD 111980 \& HD 031128 \\
 &    & 3 & 0 & 50 & 31/3/2013 & 1.00 & HD 111980 \& HD 031128 \\
 Hen2-29 & PN G275.8-02.9       &  3 & 45  & 600 & 01/4/2013 &  1.23 & HD 111980  \& HD 031128 \\
 &    & 1 & 45 & 300 & 01/4/2013 & 1.23 & HD 111980  \& HD 031128 \\
 Hen2-37  & PN G274.6+03.5    &  3  & 90 & 600 & 31/3/2013 & 1.06 &  HD 111980 \& HD 031128 \\
 \hline
 \end{tabular}}
\end{table*}

\section{Observations \& data reduction}\label{Obs}
The integral field spectra of the PNe were obtained over two nights
of March 31 and April 01, 2013 with the WiFeS instrument.
This instrument delivers a field of view of 25\arcsec $\times$
38\arcsec at a spatial resolution of either 1.0\arcsec $\times$
0.5\arcsec or 1.0\arcsec $\times$ 1.0\arcsec, depending on the
binning on the CCD. The design of the image slicer means that there
is essentially no clear space between pixels -- the spatial filling
is $\sim 98$\%.  In these observations, we operated in the binned
1.0\arcsec x 1.0\arcsec mode. The data cover the blue spectral range
of 3400-5700 {\AA} at a spectral resolution of $R \sim 3000$ that
corresponds to a full width at half maximum (FWHM) of $\sim 100$
km/s ($\sim 1.5${\AA}), while in the red spectral range of 5500-7000
{\AA} we used the higher spectral resolution grating $R \sim 7000$
corresponding to a FWHM of $\sim 45$ km/s ($\sim 0.9${\AA}).

The wavelength scale was calibrated using the Cu-Ar arc Lamp with
40s exposures throughout the night, while flux calibration was
performed using the STIS spectrophotometric standard stars HD 111980
\& HD 031128 \footnote{Available at : \newline {\tt
www.mso.anu.edu.au/$\sim$bessell/FTP/Bohlin2013/GO12813.html}}. In
addition, a B-type telluric standard HIP 38858 was observed.
Telluric absorption features from atmospheric oxygen (O$_2$) and
water (H$_2$O) are corrected with PyWiFeS \footnote
{http://www.mso.anu.edu.au/pywifes/doku.php.} data reduction
pipeline (\citet{Childress14}) as follows. First, the absorption for
each telluric standard is measured by fitting a smooth low-order
polynomial (typically a cubic function) to the stellar continuum
redward of 6000\AA, and dividing the observed spectrum by this
smooth continuum fit.  Then, the effective mean absorption at zenith
for each component (O$_2$ and H$_2$O) is computed for each observed
telluric standard by scaling the observed absorption to airmass 1
using the appropriate scaling of optical depth with airmass for each
component.  We note that the O$_2$ bands are typically saturated,
while the H$_2$O bands are not, so they have different airmass
dependences.  In addition, the relative humidity in the various
atmospheric layers ensures that the H$_2$O absorption components
vary relative to the O$_2$ obsorption. The mean zenith absorption
profiles for the two components (O$_2$ and H$_2$O) can then be
similarly scaled the airmass of any observed science field, and the
WiFeS data cube is divided by the expected telluric absorption
profile.  A similar technique can be employed with single telluric
standards whose absorption profiles are used to correct an
individual science data cube.

All data cubes were reduced using the PyWiFeS.
A summary of the spectroscopic observations is given in Table
\ref{Table1}. In some objects, due to the saturation of strong
nebular emission lines such as [O III] $\lambda$5007 and H$\alpha$,
the fluxes in these lines were derived from additional frames with
short exposure times.

The global spectra of each of the objects were extracted from  their
respective data cubes using a circular aperture matching the
observed extent of the bright region of the PNe using {\tt QFitsView
v3.1 rev.741}\footnote{{\tt QFitsView v3.1} is a FITS file viewer
using the QT widget library and was developed at the Max Planck
Institute for Extraterrestrial Physics by Thomas Ott.}. The line
fluxes measured in blue and red spectra were slightly re-scaled (by
a factor of $\sim 2\%$) using the emission lines in the overlapping
spectral range (5500-5700 {\AA}). This procedure allows for
sub-arcsec. differences in the extraction apertures caused by
differential atmospheric dispersion.

Emission-line fluxes and their uncertainties were measured, from the
final combined, flux-calibrated blue and red spectra, using the IRAF
{\tt splot} task\footnote{IRAF is distributed by the National
Optical Astronomy Observatory, which is operated by the Association
of Universities for Research in Astronomy (AURA) under a cooperative
agreement with the National Science Foundation.}. Each line was fit
with multi-gaussians, as necessary.

\section{Nebular Morphology}

\begin{figure*}
  \includegraphics[scale=0.6]{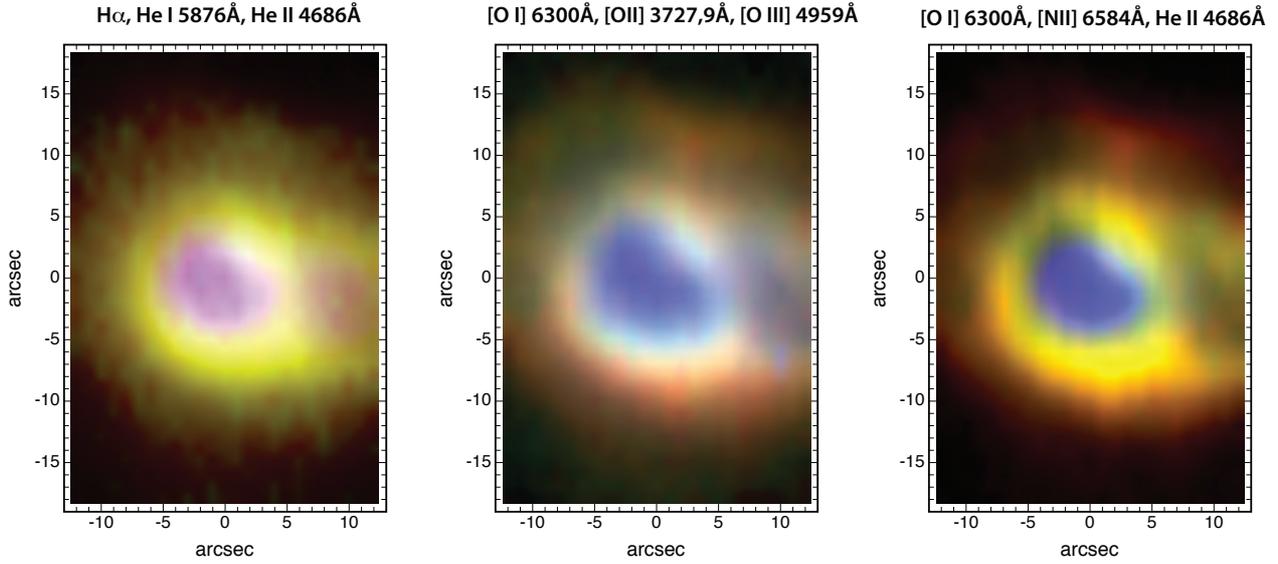}
\caption{A series of narrow-band emission-line maps for M3-4. In
this Figure, East is at the top of the image, and North is to the
right. The colour channels (R,G and B, respectively) used for the
emission line images are given above each panel. The first panel
(H$\alpha$, He\,I $\lambda 5876$ and He\,II $\lambda 4686$) gives
the clearest idea of the flux distribution in the nebula since all
three are recombination lines. The high ionisation regions appear
purple and occupy roughly the central part of the nebula, while the
lower ionisation regions appear yellow and surround the high
ionisation regions. The remaining two panels are designed to bring
out the ionisation stratification of the nebula in the oxygen ions
(central panel) and in [O\,I] $\lambda 6300$, [N\,II] $\lambda 6584$
and He\,II $\lambda 4686$ respectively (right panel). This nebula
has a complex double-shell and bipolar structure.} \label{fig1}
\end{figure*}

\begin{figure*}
  \includegraphics[scale=0.6]{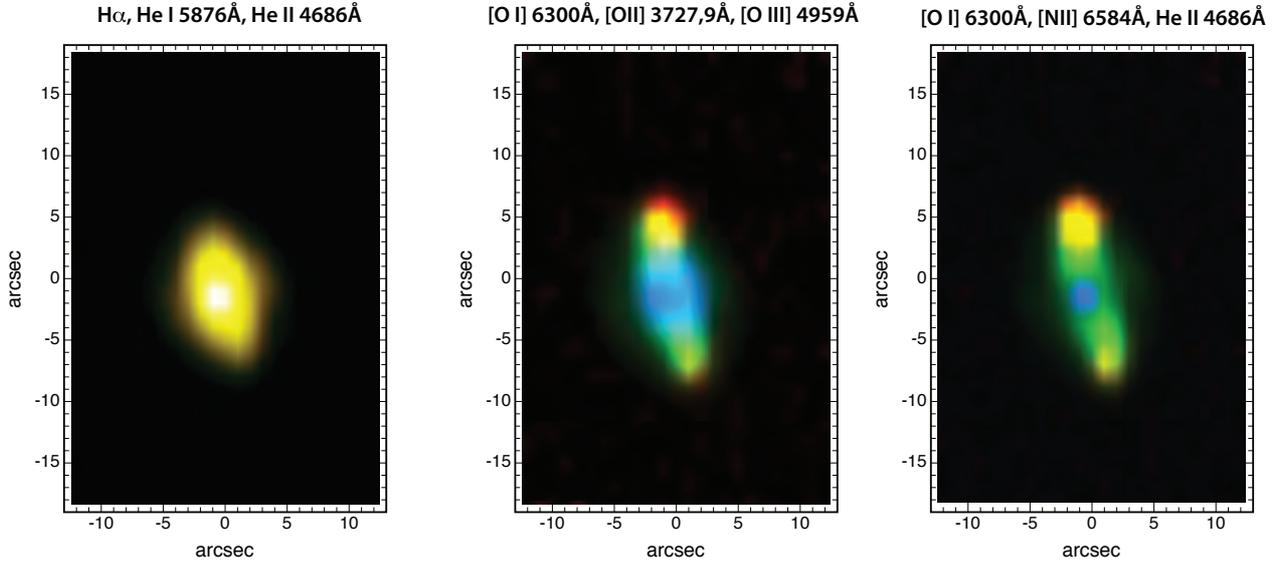}
  \caption{As Figure \ref{fig1}, but for M3-6. In this Figure, North is at the top of the image,
  and East is to the left. This object has a bright central star (visible here in its He II $\lambda 4686$ emission) with
  a strongly elliptical morphology with ``ansae''. In the excitation maps these ansae are visible in the low
  ionisation species [O I], [N II] and [O II]. The outermost knots of emission in [ O I] are probably identified as Fast low-Excitation Emission Line Regions (FLIERs). For more discussion see text.} \label{fig2}
\end{figure*}

\begin{figure*}
  \includegraphics[scale=0.6]{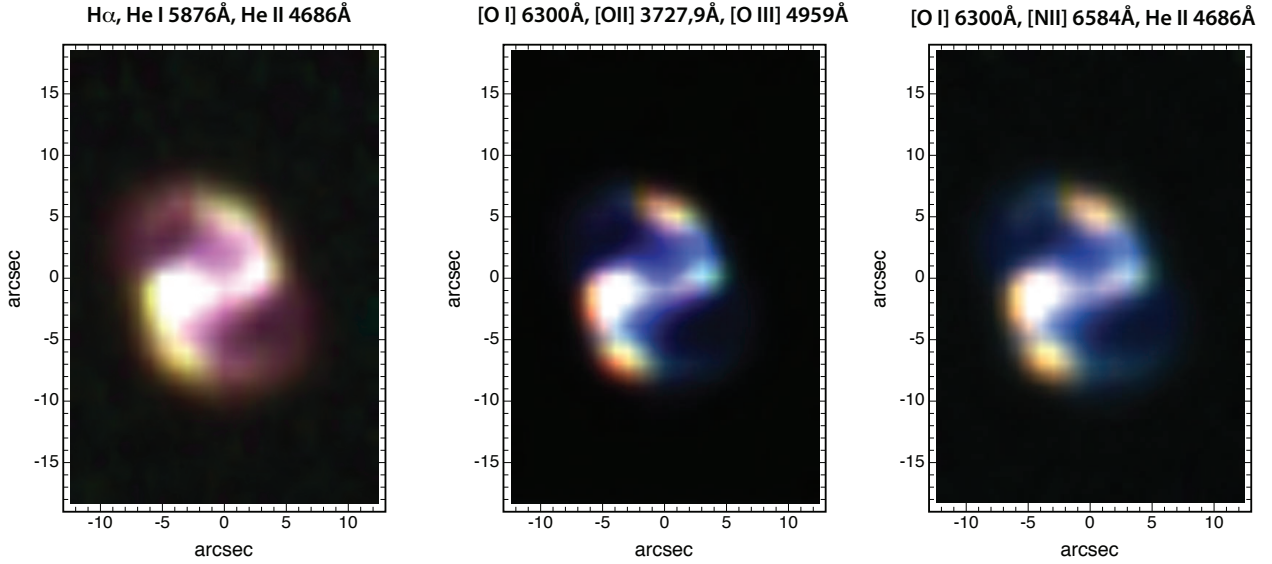}
   \caption{As Figure \ref{fig1}, but for Hen2-29. In this Figure, NE is at the top of the image, and SE is to the left. This nebula is highly bi-symmetric with four low excitation regions embedded in a two-arm barred spiral structure.} \label{fig3}
\end{figure*}

\begin{figure*}
  \includegraphics[scale=0.6]{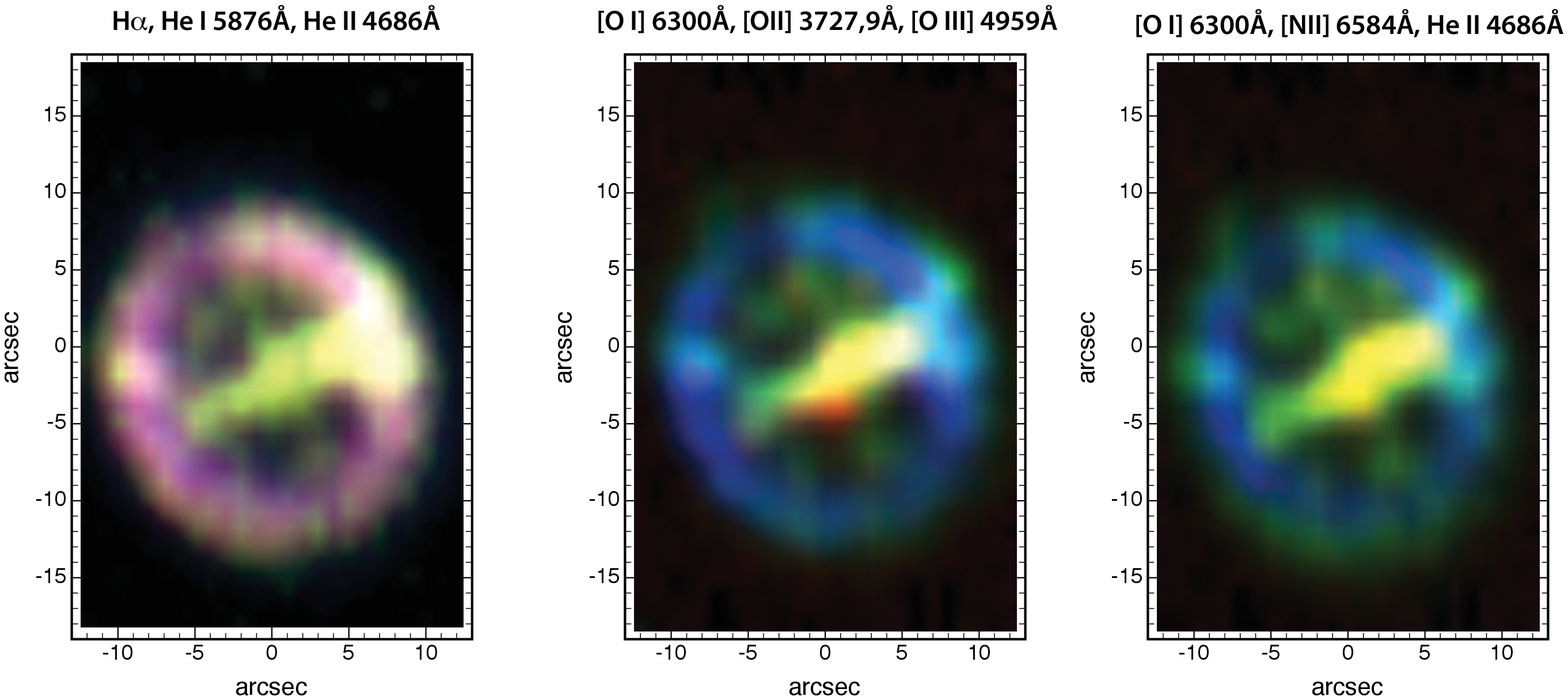}
  \caption{As Figure \ref{fig1}, but for Hen2-37. In this Figure, East is at the top of the image, and North is to the right.
  The $\Theta-$like morphology of this nebula probably arises by interaction of a symmetric fast wind with a dense equatorial
  disk of material ejected during the AGB phase of evolution. This region also contains a number of denser knots
  undergoing photo-evaporation.} \label{fig4}
\end{figure*}

Emission-Line maps of the brightest lines were constructed using the
reduced  WiFeS data cubes for each nebula in our set. These maps are
used to describe their internal flux distributions and their
excitation structure, by combining three lines into an RGB image. We
have selected three such images for inclusion here, and these are
presented for each nebula in Figures \ref{fig1} to \ref{fig4}.

First, a combination of H$\alpha$, He\,I $\lambda 5876$ and He\,II
$\lambda 4686$ (left panels in Figures \ref{fig1} to \ref{fig4})
provides  the emissivity distribution in the principal recombination
lines. Since H$\alpha$ is distributed throughout the nebula, in
regions where helium is not ionised, these maps will be dominated by
the H$\alpha$ flux, and will appear red. In regions where Helium is
singly ionised, the map will appear yellow (R+G), while in regions
in which the He II $\lambda 4686$ line is seen, the map will appear
mauve (R+B).

The second combination of the lines [O\,I] $\lambda 6300$, [O\,II]
$\lambda 3727,9$ and [O\,III] $\lambda 5007$ -- the central panels
in Figures \ref{fig1} to \ref{fig4} -- brings out the ionisation
stratification very clearly. Finally, as a different way of bringing
out the ionisation structure, we give the [O\,I] $\lambda 6300$,
[N\,II] $\lambda 6584$ and He\,II $\lambda 4686$ image (right panels
in Figures \ref{fig1} to \ref{fig4}). This combination covers a more
extreme range in ionisation potential than the central panels.

These four PNe have very different morphologies. M3-4 (Figure
\ref{fig1}) has a very complex bipolar double-shell structure with
two lobes in which He\,II $\lambda 4686$ is strong. The appearance
of the inner bright rings suggest that they represent an incomplete
bipolar shell. The ionisation stratification across these rings is
clearly evident. This double-ring structure is embedded in an outer
shell dominated by the [O\,I] $\lambda 6300$ emission. The outer
shell has similar morphology to the inner, but it extends beyond the
field of view of this WiFeS frame. The green tones in the upper left
and right quadrants of the central panel shows that the outer shell
is of intermediate excitation with strong [O\,II] $\lambda 3727,9$
emission.

M3-6 (Figure \ref{fig2}) appears as elliptical PN with brightness
increasing towards the centre. Overall, the nebula is of lower
excitation than M3-4, and the He II $\lambda 4686$ emission has its
origin in the central star. It shows pronounced ``ansae'' along
PA$\sim 20\deg$. At the ends of these ansae are two bright knots
very prominent in [O I] $\lambda 6300$ and [S II] $\lambda 6717,31$.
These are almost certainly Fast Low-Excitation Emission Line Regions
(FLIERs; \citet{Balick87, Balick93, Balick94}). These regions are
almost certainly shock-excited by a convergent fast stellar wind
flow shaped by the density and/or magnetic structure of the earlier
AGB wind \citep{Dopita97, Steffen02}. However, the radial velocities
of these two knots are very little different -- we measure the
northern knot to have a heliocentric radial velocity of
$34.7\pm3.4$\,km\,s$^{-1}$ and the southern one at
$23.8\pm3.3$\,km\,s$^{-1}$. If our hypothesis on their origin is
correct, the knots must lie almost in the plane of the sky. In Table
7, we determine an overall heliocentric velocity of $31.1\pm3.5$ for
the nebula which supports our interpretation.

At very low flux levels, we also detected a faint extended halo
surrounding the bright nebula and extending over a region of $\sim$
14\arcsec x 20\arcsec. This may be the ionised trace of the earlier
AGB wind.

Hen2-29, Figure \ref{fig3} reveals "a reverse S-shape" or two-arm
spiral structure formed by two point-symmetric arcs present along
the minor axis within an overall elliptical morphology. The densest
regions in the arms are also optically thick, while the overall
elliptical outline is filled with high-excitation gas.

Finally, Hen2-37 (Figure \ref{fig4}) displays a $\Theta -$like
morphology overall. The bar of the $\Theta$ contains optically thick
material, and there are isolated lower recitation knots embedded in
the body of the nebula in a region of otherwise low density. The
high excitation regions are confined to a region within the outer
circle of the  $\Theta$, and is bounded at its outer boundary by
lower excitation gas. Here, we might interpret the bar of the
$\Theta$ as a remnant of a dense equatorial ring of martial ejected
during the AGB phase, the knots as being photevaporating inclusions
embedded in the fast stellar wind region, and the outer ring as
being the swept-up shell of denser AGB wind gas.

\section{Determination of physical conditions and abundances}

\subsection{Line intensities and reddening corrections}
The global emission line spectra are summarised in Table
\ref{Table2}. These have been analysed using the Nebular Empirical
Abundance Tool (NEAT \footnote{The code, documentation, and atomic
data are freely available at
http://www.nebulousresearch.org/codes/neat/.}; \citet{Wesson12}).
The code was applied to derive the interstellar extinction
coefficient and subsequent nebular parameters such as temperature,
density and ionic and elemental abundances. The line intensities
have been corrected for extinction by adopting the extinction law of
\citet{Howarth83}, where the amount of interstellar extinction was
determined from the ratios of hydrogen Balmer lines, in an iterative
method. Firstly, an initial value for the extinction coefficient,
$c(H\beta)$, was calculated by the code. This value was derived from
the intrinsic $H\alpha$, $H\beta$, $H\gamma$ line ratios, assuming
an electron temperature ($T_e$) of 10000K and electron density
($N_e$) of 1000 cm$^{-3}$. Secondly, the code calculates the
electron temperature and density as explained in the coming section.
Thirdly, the intrinsic Balmer line ratios were re-calculated at the
appropriate temperature and density, and again $c(H\beta)$ was
recalculated. No spatial reddening variation was detected throughout
any of the PN set, except in the case of M3-6. The two
low-ionization condensations which lie along its major axis show a
lower extinction compared to the main nebula.

Table \ref{Table3}, lists the estimated reddening coefficients and
the observed $H\beta$ and $H\alpha$ fluxes on a log scale. The
complete list of the observed and de-reddened line strengths are
given  in Table \ref{Table2}. Here, columns (1) and (2) give the
laboratory wavelengths ($\lambda_{\rm Lab}$) and the line
identifications of observed emission lines respectively. Columns
3-14 give the observed wavelength ($\lambda_{\rm Obs}$), fluxes
F($\lambda$) and fluxes corrected for reddening I($\lambda$)
relative to $H\beta =100$.  The Monte Carlo technique was used by
NEAT to propagate the statistical uncertainties from the line flux
measurements through to the derived abundances. Both line ratios
[O\,III] $\lambda$5007/$\lambda$4959 and [N\,II]
$\lambda$6584/$\lambda$6548 were measured to be in the range of
(2.9-3.1) which are comparable to the theoretical predictions of
2.98 \citep{Storey00} and 2.94 \citep{Mendoza83}, respectively.

\begin{table*}
 \caption{Integrated line fluxes and derived the dereddened line intensities for M3-4, M3-6, Hen2-29 and Hen2-29. Absolute line fluxes for H$\beta$ are given
 in the text. The C IV doublet lines 5801+5812\AA \, are of CS origin.The full
version of the table is available online. A portion is shown here
for guiding the reader regarding its content.} \label{Table2}
\scalebox{0.65}{
\begin{tabular}{llcccccccccccc}
 \hline
 &  & \multicolumn{3}{c}{M3-4} & \multicolumn{3}{c}{M3-6}  & \multicolumn{3}{c}{Hen2-29} & \multicolumn{3}{c}{Hen2-37}\\
$\lambda_{\rm Lab}$ (\AA)    & ID   & $\lambda_{\rm Obs}$ (\AA) &
F($\lambda$) &
I($\lambda$) & $\lambda_{\rm Obs}$ (\AA) & F($\lambda$) & I($\lambda$) & $\lambda_{\rm Obs}$ (\AA) & F($\lambda$) & I($\lambda$) & $\lambda_{\rm Obs}$ (\AA) & F($\lambda$) & I($\lambda$)\\
\hline
4685.68 &   He   II &   4686.29 &   33.2$\pm$0.99   &   34.0$\pm$1.51   &   4686.47 &   0.2$\pm$0.02    &   0.2$\pm$0.02    &   4685.20 &   65.6$\pm$3.28   &   71.7$\pm$5.35   &   4685.20 &   67.9$\pm$3.40   &   73.3$\pm$5.48    \\
4711.37 &   [Ar  IV]    &   4712.13 &   2.2$\pm$0.34    &   2.3$\pm$0.36    &   4711.56 &   0.3$\pm$0.11    &   0.3$^{+0.06}_{-0.08}$   &   4710.86 &   3.7$\pm$0.39    &   4.0$\pm$0.48    &   4710.80 &   3.4$\pm$0.32    &   3.6$\pm$0.40    \\
4714.17 &   [Ne  IV]    &       &       &       &       &       &       &   4713.34 &   0.8$\pm$0.17    &   0.7$^{+0.132}_{-0.163}$ &   4713.55 &   0.3$\pm$0.06    &   0.3 $^{  +0.054}_{  -0.066}$\\
4724.89 &   [Ne  IV]b   &       &       &       &       &       &       &   4724.20 &   0.8$\pm$0.12    &   0.9$\pm$0.14    &   4724.20 &   0.6$\pm$0.09    &   0.6$\pm$0.10    \\
4740.17 &   [Ar  IV]    &   4740.76 &   1.7$\pm$0.25    &   1.7$\pm$0.26    &   4740.52 &   0.4$\pm$0.05    &   0.4$\pm$0.06    &   4739.70 &   2.9$\pm$0.29    &   3.1$\pm$0.35    &   4739.50 &   2.3$\pm$0.23    &   2.5$\pm$0.28    \\
4861.33 &   HI   4-2    &   4861.79 &   100.0$\pm$3.00  &   100.0$\pm$0.00  &   4861.55 &   100.0$\pm$5.00  &   100.0$\pm$0.00  &   4860.70 &   100.0 $\pm$5.00 &   100.0$\pm$0.00  &   4860.60 &   100.0$\pm$5.00  &   100.0$\pm$0.00    \\
4921.93 &   He   I  &   4922.8  &   0.9$\pm$0.14    &   0.9$\pm$0.14    &   4922.11 &   1.2$\pm$0.19    &   1.2$\pm$0.20    &   4921.30 &   1.2$\pm$0.12    &   1.2$\pm$0.13    &   4921.20 &   0.9$\pm$0.13    &   0.9$\pm$0.14    \\
4958.91 &   [O   III]   &   4959.38 &   403.3$\pm$12.10 &   398.2$\pm$16.63 &   4959.11 &   232.2$\pm$11.61 &   227.3$\pm$15.95 &   4958.30 &   566.6 $\pm$28.3 &   540.8$\pm$37.42 &   4958.20 &   722.8$\pm$36.14 &   695.5$\pm$48.36   \\
5006.84 &   [O   III]   &   5007.28 &   1198.9$\pm$36.0 &   1175.6$\pm$48.9 &   5007.02 &   695.4$\pm$34.8  &   673.1$\pm$47.0  &   5006.20 &   1696.3$\pm$84.8 &   1582.4$\pm$107.3    &   5006.20 &   2238.3$\pm$111.9    &   2110.7$\pm$145.2  \\
\hline
\end{tabular}}
\end{table*}

\subsection{Optical thickness estimates}
The low ionisation [N\,I] doublet $\lambda5197$ and $\lambda5201$ is
present in the spectra of M3-4, Hen2-29 and Hen2-37. It is known
that this line and other low ionization species such as [O\,I]
$\lambda6300,6363$ arise in the outermost part of the nebula. Due to
the charge-transfer reaction rate, it is expected that both [O\,I]
and [N\,I] are abundant in the warm transition region between the
neutral and ionized nebular envelopes (Liu, X-W et al. 1995). If
shocks are present, the intensity of these lines is increased thanks
to the increase in collision strength with electron temperature. In
the case of M3-4 nebula, both [N\,I]  and [O\,I] are present in the
external parts of the object. By contrast, in Hen2-37 we found
traces of [N\,I]  and [O\,I] in the central region of the object
(see Figure \ref{fig4}, central panel), showing that the bar of the
$\Theta$ is composed of material which is optically-thick to the
radiation field. In the case of Hen2-29, the [N\,I] gas appears
distributed in several regions of high emissivity (see Figure
\ref{fig3}). Using [N\,I] line ratio $\lambda5197$/$\lambda5201$, we
estimate ionic densities of 378cm$^{-3}$ and 412cm$^{-3}$ for M3-4
and Hen2-29, respectively. These estimates are in good agreement
with the other nebular density measurements listed in Table
\ref{Table4}.

In the case of Hen2-37, we can crudely estimate an electron
temperature, $T_e = 14700$K, from the [O\,I] $\lambda\lambda
5577/6300$ line ratio for Hen2-37  nebula. This value is higher than
the temperatures derived from [O\,III] and [N\,II] lines (Table
\ref{Table4}), and may indicate the presence of a shock
contribution.

The strength of [O\,II] and [N\,II] emission lines were used by
\citet{Kaler89} to help define the optical thickness of PNe.
Following their criteria, we verified that Hen2-37 nebula is
probably optically thick to the ionising radiation field (radiation
bounded). It displays strong low excitation lines, with F([N\,II]
$\lambda$6583)$\geq$F(H$\alpha$) and F([O\,II]
$\lambda3727$)/F(H$\beta$) $\geq$ 1.5. By contrast, the nebula M 3-6
is optically thin (density bounded), except within the ansae. This
is supported by several lines of evidence: (1) The weakness of the
low excitation lines [O\,I], [O\,II] and [N\,II]; (2) A line ratio
F([N\,II]$\lambda6583$)/F(H$\alpha) \leq 0.1$; (3) The absence of
detectable [N\,I] at $\lambda$ 5200. The other two nebulae in the
sample appear partially optically thick PNe. M3-4 and Hen2-29 have
F[N\,II]($\lambda6583$)/F(H$\alpha)= 0.9$,
F[O\,II]($\lambda$3727)$/$F(H$\beta)=1.25$,
F[N\,II]($\lambda$6583)$/ $F(H$\alpha)= 0.75$ and
F[O\,II]($\lambda$3727)$/$F(H$\beta)=1.56$, respectively.

\begin{table*}
\centering \caption{Reddening coefficients and observed $H\beta$ and
$H\alpha$ fluxes of PNe set.} \label{Table3} \scalebox{0.9}{
\begin{tabular}{lllllll}
 \hline
 Object & \multicolumn{2}{|c|}{$c(H\beta)$} & \multicolumn{2}{|c|}{Log F$(H\beta)$}  & \multicolumn{2}{|c|}{Log F$(H\alpha)$} \\
        & This work & Other works & This work & Other works & This work & Other works \\
 \hline
M3-4   & 0.25     & 0.2$^1$, 0.38$^2$, 0.70$^3$, 0.23$^4$   &-12.43 & -12.4$^7$, -12.60$^2$  & -11.49 & -11.46$^9$ \\
M3-6  & 0.43     & 0.64$^1$, 0.35$^2$, 0.58$^3$, 0.49$^5$, 0.53$^6$  & -11.01 & -11.1$^7$, -11.22$^2$, -12.0$^5$ & -10.41 & -10.36$^9$   \\
Hen2-29 & 0.87     & 1.01$^1$, 0.63$^3$   & -12.17  & -12.1$^7$, -12.17$^8$ & -11.44 & -11.40$^9$ \\
Hen2-37  & 0.74    &  0.80$^1$, 0.69$^2$, 1.03$^3$   & -12.28 &-12.4$^7$, -12.73$^2$, -12.40$^8$  & -11.60 & -11.53$^9$\\
 \hline
\end{tabular}}
\begin{minipage}[!t]{15cm}
{\tiny {\bf References.}: (1) \citet{Tylenda92};(2)
\citet{Milingo02}; (3) \citet{Kingsburgh94};(4) \citet{Kniazev12}; (5) \citet{Girard07};
(6)  \citet{de-Freitas92}; (7) \citet{Acker91};  (8) \citet{Perinotto04}; (9) \citet{Frew13}}. \\
\end{minipage}
\end{table*}

\begin{table*}
\centering \caption{Electron temperatures and densities of the PNe
set. } \label{Table4} \scalebox{0.85}{
\begin{tabular}{lllllllll}
 \hline
      Object         & \multicolumn{4}{|c|}{Temperature (K)} & \multicolumn{4}{|c|}{Density (cm$^{-3}$)} \\
               & {[}O III{]} & {[}N II{]} &  {[}S II{]} & {[}Ar V{]} & {[}S II{]} & {[}O II{]} & {[}Ar IV{]} & {[}Cl III{]} \\
                \hline
M3-4         & 11154$^{+396}_{-396}$     & 10267$^{+227}_{-227}$   & & & 362$^{+132}_{-229}$     & 721$^{+185}_{-147}$     &  285$^{+166}_{-220}$   & 1202$^{+720}_{-516}$  \\
Ref (1)         &  11500   & 9300  & 8500   &       &  200      &      &  & \\
Ref (2) & 12100$^{+1000}_{-1000}$     & 9900  &    &      & 100$^+$
& \\
Ref (3) & 12103$^{+360}_{-360}$     &   & & & 322$^{+142}_{-142}$ & & \\

\hline
M3-6        & 7988$^{+212}_{-212}$     & 10156$^{+485}_{-485}$   & & & 2568$^{+685}_{-541}$     & 4289$^{+1768}_{-4289}$ &  2196$^{+725}_{-593}$ & 2013$^{+443}_{-605}$  \\
Ref (1) & 8000     & 8700  &   &      &   1700     &       &   &  \\
Ref (2) &  8800:    &   &    &      &        & 3020$^{+370}_{-330}$      &   \\
Ref (4) & 8100 & 11300  &    &      &  5210      &       &   \\
Ref (5) & 9000 &   &    &      &        &       &   \\

\hline
Hen2-29       & 14329$^{+420}_{-420}$     & 12520$^{+272}_{-272}$    & 12038$^{+1542}_{-12038}$ &  & 855$^{+137}_{-137}$     & 1075$^{+373}_{-277}$  & 351$^{+158}_{-88}$  & 423$^{+169}_{-166}$ \\
Ref (2) & 17100$^{+500}_{-600}$      & 11000$^{+3500}_{-2500}$  &    &      & 720$^{+300}_{-240}$      &       &   \\

\hline
Hen2-37       & 12147$^{+300}_{-300}$     & 11324$^{+397}_{-397}$    & 10743$^{+1231}_{-1104}$ & & 361$^{+93}_{-93}$     & 697$^{+325}_{-193}$ &  209$^{+125}_{-83}$ & 341$^{+123}_{-188}$ \\
Ref (1) & 12000&  10000 & 10300 &  & 200 &  &  \\
Ref (2) & 12600$^{+400}_{-200}$ & 10100$^{+200}_{-500}$ & &
 &190$^{+30}_{-30}$ & & 1600$^{+280}_{-240}$ & \\
Ref (6) &  12824    & 10169  &    &      & 270       &       &   \\
\hline
\end{tabular}}
\begin{minipage}[!t]{16cm}
{\tiny References: (1) \citet{Milingo02}; (2) \citet{Kingsburgh94};
(3) \citet{Kniazev12}; (4) \citet{Girard07}; (5)
\citet{de-Freitas92}; (6) \citet{Martins00}. The symbol (+)
indicates the diagnostic ratio is at low density limit, (:)
indicates $\sim 50$ per cent uncertainty in the measured flux.}\\
\end{minipage}
\end{table*}

\subsection{Temperatures and densities}

The observed emission lines in our sample (Table \ref{Table2}),
permitted  us to estimate the electron temperature and density from
both the low and the medium ionisation zones. The low ionisation
species provide the nebular temperature from the [N\,II] ($\lambda
6548 + \lambda 6584$)/$\lambda$5754 line ratio and the density from
the [S\,II] $\lambda$6716/$\lambda$6731 and the [O\,II]
$\lambda$3727/$\lambda$3729 line ratios.  The moderately excited
species provide the nebular temperature from the [O\,III] ($\lambda
4959 + \lambda 5007$)/$\lambda$4363 line ratio and the density from
the [Cl\,III] $\lambda$5517/$\lambda$5537 and [Ar\,IV]
$\lambda$4711/$\lambda$4740 line ratios. The temperatures derived
from [O\,III] lines are consistently higher than those derived from
the [N\,II] lines, except in the case of the optically-thin object
M3-6.

 In Table \ref{Table4}, we list the calculated nebular temperatures,
 densities and their uncertainties for each object. In addition, it
 also provides comparisons of our results with those which have previously
 appeared in the literature. In almost cases, we find good agreement with
 other works. As expected, temperatures derived from [O\,III] lines are
 consistently higher than those derived from the [N\,II] lines, except
 in the case of the optically-thin object M3-6.

\subsection{Ionic and elemental abundances}
Applying the NEAT, ionic abundances of Nitrogen, Oxygen, Neon,
Argon, Chlorine and Sulfur were calculated from the collisional
excitated lines (CEL), while Helium and Carbon were calculated from
the optical recombination lines (ORL) using the temperature and
density appropriate to their ionization potential. When several
lines from a given ion are present, the adopted ionic abundance  was
taken averaging the abundances from each line weighted according to
their line intensities. The total elemental abundances were
calculated from ionic abundances using the ionization correction
factors (ICF) given by \citet{Kingsburgh94} to correct for unseen
ions. The helium elemental abundances for all objects were
determined from the He$^+$/H and He$^{2+}$/H ions, assuming ICF (He)
= 1.0. The carbon elemental abundances were determined from
C$^{2+}$/H for all objects, except in the case of Hen2-29 where we
used both C$^{2+}$/H and C$^{3+}$/H taking ICF (C) = 1.0. In
general, the chemical abundances are found to be in good agreement
with other works which presented in Tables \ref{Table5} and
\ref{Table6}.

\begin{table*}
\centering \caption{Abundances derived from the NEAT for M3-4 and
M3-6 compared with other works. Neat (1) and Neat (2) refer to the
derived abundances using the ionization correction factors of
\citet{Kingsburgh94} and \citet{Delgado14}, respectively.}
\label{Table5} \scalebox{0.75}{
\begin{tabular}{lcccccccccccc}
\hline
    & \multicolumn{6}{|c|}{M 3-4} & \multicolumn{6}{|c|}{M 3-6} \\
\hline
Element & Neat(1) & Neat (2) & Ref 1 & Ref 2  & Ref 3 & Ref 4  & Neat(1) & Neat (2) & Ref 1 & Ref 2  & Ref 3 & Ref 4 \\
\hline He/H & 1.15E-1$^{+5.3{\rm E}-3}_{-5.3{\rm E}-3}$ &
1.15E-1$^{+4.3{\rm E}-3}_{-4.3{\rm E}-3}$ & 1.23E-1 &
--- & 1.5E-1 & --- & 1.11E-1$^{+5.1{\rm E}-3}_{-5.1{\rm E}-3}$ & 1.11E-1$^{+3.7{\rm E}-3}_{-3.7{\rm E}-3}$ &
1.20E-1
& 0.90E-1 & --- & --- \\
C/H & 6.08E-4$^{+9.4{\rm E}-5}_{-9.4{\rm E}-5}$ & 5.86E-4$^{+1.4{\rm
E}-4}_{-8.1{\rm E}-5}$ & --- & --- & --- &
--- & 3.68E-4$^{+5.5{\rm E}-5}_{-5.5{\rm E}-5}$ & 7.80E-4$^{+5.5{\rm E}-5}_{-5.5{\rm E}-5}$ & --- & ---
 & --- & --- \\
N/H &  2.10E-4$^{+2.4{\rm E}-5}_{-2.2{\rm E}-5}$ &
2.10E-4$^{+3.3{\rm E}-5}_{-2.9{\rm E}-5}$ & 1.6E-4 & --- & 1.93E-4 &
1.74E-4 & 6.83E-5$^{+1.6{\rm E}-5}_{-1.3{\rm E}-5}$ &
1.37E-5$^{+3.1{\rm E}-5}_{-3.5{\rm E}-5}$ & 0.58E-4 & 0.24E-4
& --- &  0.23E-4\\
O/H & 4.55E-4$^{+4.7{\rm E}-5}_{-4.2{\rm E}-5}$ & 3.97E-4$^{+4.8{\rm
E}-5}_{-4.8{\rm E}-5}$ & 5.22E-4 & 5.29E-4 & 5.16E-4 & 3.80E-4 &
5.55E-4$^{+7.9{\rm E}-5}_{-6.9{\rm E}-5}$  & 5.24E-4$^{+6.2{\rm
E}-5}_{-5.3{\rm E}-5}$ & 5.59E-4 &
2.8E-4 & 4.4E-4 & 4.4E-4 \\
Ne/H & 1.48E-4$^{+2.2{\rm E}-5}_{-1.9{\rm E}-5}$  &
1.47E-4$^{+1.7{\rm E}-5}_{-1.4{\rm E}-5}$ & 1.5E-4 & 1.5E-4 &
1.76E-4 & --- & 1.45E-4$^{+2.3{\rm E}-5}_{-2.0{\rm E}-5}$ &
1.58E-4$^{+2.1{\rm E}-5}_{-2.1{\rm E}-5}$ & 1.36E-4 &
0.86E-4 & 5.13E-5 & --- \\
Ar/H & 6.27E-7$^{+1.1{\rm E}-7}_{-9.2{\rm E}-8}$ &
6.27E-7$^{+1.1{\rm E}-7}_{-9.2{\rm E}-8}$ & ---  & 2.82E-6 & 1.43E-6
& ---
 & 2.40E-7$^{+4.6{\rm E}-8}_{-3.9{\rm E}-8}$ & 2.40E-7$^{+4.6{\rm E}-8}_{-3.9{\rm E}-8}$ & 3.06E-7 &  ---
 & 2.09E-6 & --- \\
S/H & 2.28E-6$^{+4.7{\rm E}-7}_{-3.9{\rm E}-7}$ & 2.50E-6$^{+5.3{\rm
E}-7}_{-6.2{\rm E}-7}$ & --- & 3.1E-6 & 0.16E-5 &
--- & 1.20E-5$^{+3.1{\rm E}-6}_{-2.4{\rm E}-6}$ &
1.01E-5$^{+2.2{\rm E}-6}_{-1.1{\rm E}-6}$ & 0.71E-5 &
--- & 5.37E-6 & --- \\
Cl/H & 5.45E-8$^{+7.7{\rm E}-9}_{-6.8{\rm E}-9}$  &
5.79E-8$^{+8.5{\rm E}-9}_{-9.2{\rm E}-9}$ & --- & --- & 0.84E-7 &
--- & 2.83E-7$^{+5.3{\rm E}-8}_{-7.0{\rm E}-8}$ & 2.83E-7$^{+5.3{\rm E}-8}_{-7.0{\rm E}-8}$ & 1.18E-7 & --- &
--- &  --- \\
N/O & 0.46   & 0.52    &    ---   & ---    &   0.37 & ---
& 0.12  &  0.26 & 0.10 & ---   & --- & --- \\
\hline
\end{tabular}}
\begin{minipage}[!t]{15cm}
{\tiny References: (1) \citet{Milingo02}; (2) \citet{Kingsburgh94};
(3) \citet{Maciel99};
(4) \citet{Chiappini94}; (5) \citet{Perinotto04}; (6) \citet{Martins00}.}\\
\end{minipage}
\end{table*}

\begin{table*}
\centering \caption{Abundances derived from the NEAT for Hen2-29 and
Hen2-37  compared with other works. Neat (1) and Neat (2) refer to
the derived abundances using the ionization correction factors of
\citet{Kingsburgh94} and \citet{Delgado14}, respectively.}
 \label{Table6}
\scalebox{0.75}{
\begin{tabular}{lcccccccccccccc}
 \hline
  &   \multicolumn{5}{|c|}{Hen2-29 } & \multicolumn{7}{|c|}{Hen2-37 } \\
\hline
Element & Neat(1) & Neat (2) & Ref 1 & Ref 2 & Ref 3 & Neat(1) & Neat (2) & Ref 1 & Ref 2  & Ref 3 & Ref 5 & Ref 6 \\
\hline He/H & 1.26E-1$^{+7.3{\rm E}-3}_{-7.3{\rm E}-3}$ &
1.26E-1$^{+4.9{\rm E}-3}_{-4.9{\rm E}-3}$ & 1.09E-1 & 1.08E-1 & ---
& 1.17E-1$^{+4.0{\rm E}-3}_{-4.0{\rm E}-3}$ & 1.17E-1$^{+4.0{\rm E}-3}_{-4.0{\rm E}-3}$ & 1.20E-1 & 1.19E-1  & --- & 1.04E-1 & ---\\

C/H & 1.63E-3$^{+2.2{\rm E}-4}_{-2.2{\rm E}-4}$ & 1.63E-3$^{+2.2{\rm
E}-4}_{-2.2{\rm E}-4}$& --- & --- & --- &7.56E-4$^{+1.2{\rm
E}-4}_{-1.1{\rm E}-4}$ &7.56E-4$^{+1.3{\rm E}-4}_{-1.3{\rm E}-4}$ &
--- & --- & --- & ---
& --- \\
 N/H & 2.17E-4$^{+2.5{\rm E}-5}_{-3.6{\rm E}-5}$ & 1.80E-4$^{+2.1{\rm E}-5}_{-2.5{\rm E}-5}$ &  1.39E-4 &
1.66E-4 & --- & 3.18E-4$^{+5.0{\rm E}-5}_{-3.8{\rm E}-5}$  &
2.40E-4$^{+3.1{\rm E}-5}_{-4.2{\rm E}-5}$ & 3.52E-4 & 2.36E-4
 & --- & 2.20E-4 & 3.89E-4\\

O/H & 3.77E-4$^{+3.8{\rm E}-5}_{-3.1{\rm E}-5}$ & 3.79E-4$^{+3.9{\rm
E}-5}_{-3.2{\rm E}-5}$ & 4.04E-4 & 4.15E-4 & 4.17E-4 &
7.08E-4$^{+6.4{\rm E}-5}_{-7.5{\rm E}-5}$ & 7.23E-4$^{+6.7{\rm
E}-5}_{-7.1{\rm E}-5}$ & 11.0E-4 & 8.42E-4
 & 9.12E-4 & 9.34E-4 & 11.2E-4\\

Ne/H &1.01E-4$^{+8.4{\rm E}-5}_{-8.5{\rm E}-5}$ &8.67E-5$^{+7.7{\rm
E}-6}_{-8.2{\rm E}-6}$ & 1.09E-4 & 4.81E-5 & 5.01E-5 &
1.52E-4$^{+1.4{\rm E}-5}_{-1.3{\rm E}-5}$ & 1.41E-4$^{+1.2{\rm
E}-5}_{-1.7{\rm E}-5}$ & 2.48E-4 & 2.0E-4
  & 1.73E-4 &  2.30E-4 & 1.02E-4\\
Ar/H
 & 1.55E-6$^{+2.3{\rm E}-7}_{-2.3{\rm E}-7}$ & 1.96E-6$^{+4.0{\rm E}-7}_{-3.8{\rm E}-7}$ & 1.27E-6 & 1.37E-6 & 1.38E-6
 & 2.10E-6$^{+3.9{\rm E}-7}_{-2.7{\rm E}-7}$ & 3.17E-6$^{+6.1{\rm E}-7}_{-8.6{\rm E}-7}$ & 4.75E-6 &  2.91E-6
 & ---  &  2.86E-6 & ---\\
S/H & 4.95E-6$^{+5.5{\rm E}-7}_{-6.6{\rm E}-7}$ & 6.31E-6$^{+7.1{\rm
E}-7}_{-9.8{\rm E}-7}$ & 6.44E-6 & 9.07E-6 & 9.12E-6 &
8.51E-6$^{+1.1{\rm E}-6}_{-1.2{\rm E}-6}$ & 1.20E-5$^{+2.2{\rm
E}-6}_{-1.6{\rm E}-6}$ & 4.5E-6 & 6.73E-6
 & ---  & 6.60E-6 & 1.35E-5\\
Cl/H & 1.14E-7$^{+1.1{\rm E}-8}_{-1.1{\rm E}-8}$ &
1.39E-7$^{+1.5{\rm E}-8}_{-1.7{\rm E}-8}$ & -- & --- & --- &
2.08E-7$^{+2.2{\rm E}-8}_{-2.3{\rm E}-8}$ & 2.73E-7$^{+2.8{\rm
E}-8}_{-3.3{\rm E}-8}$ & 2.98E-7 & ---
& ---  & --- & ---\\
N/O & 0.54  &  0.48&  0.34       &     ---   &    ---
& 0.40  &  0.33 &  0.32 &   --- &  0.24  & --- \\
 \hline
\end{tabular}}
\begin{minipage}[!t]{14cm}
{\tiny References: (1) \citet{Perinotto04}; (2)
\citet{Kingsburgh94}; (3) \citet{Maciel99}; (4) \citet{Milingo02};
(5) \citet{Chiappini94}.}
\end{minipage}
\end{table*}

\subsection{Excitation classes and Peimbert classification }
Our objects (Table \ref{Table2}), display a mixture of both low- and
high-excitation emission lines.  Furthermore, the highly ionised
species [Ne\,V] appears in M3-4, Hen2-29  and Hen2-37. The line
strength of He II $\lambda$4686{\AA} relative to $H\beta$ provides a
best quantitative measure of the nebular excitation class (EC). The
absence of the He II line from the nebular spectrum restricts the
PNe to be of low excitation class (EC $\textless$ 5), and where it
appears (EC $\geq$ 5), its strength relative to H$\beta$ can be used
to define the excitation class. To derive the EC of our PN set, we
were followed the methodology of \citet{Meatheringham91} and
\citet{Reid10}. Both methods use the same scheme to derive the EC of
low excitation PNe, and therefore give the same result. For PNe of
EC$\geq 5$, \citet{Reid10} have upgrade the scheme of
\citet{Meatheringham91} by considering also  the increase of
[O\,III]$/H\beta$ line ratio with nebular excitation. This line
ratio was fixed in the scheme of \citet{Meatheringham91}, since in
their optically-thick models this line ratio `saturates'. However,
many PNe are optically thin, and in this case, the [O\,III]$/H\beta$
line ratio initially increases with decreasing optical thickness,
before finally decreasing for high-excitation objects when the
nebula becomes too optically-thin. Our results (Table \ref{Table6})
reveal that M3-6 is of low EC, while other three objects are of
medium to high EC.

Hen2-29  is the only PN in our set which shows excess in He and N
abundances. Thus, it can be classified as a Peimbert type I by
applying the two criteria, He/H $\geq 0.125$ and N/O $\geq 0.5$, as
proposed by \citet{Maciel99}. These criteria are much rigid than the
original criteria, He/H $\geq 0.125$ or N/O $\geq 0.5$, suggested by
\citet{Peimbert83}. Based on the analysis of 68 PNe,
\citet{Kingsburgh94} found that the average helium abundance of Type
I PNe is increased by a small amount (a factor of 1.2) over non-Type
I objects. Therefore, they distinguish between Type I and non-Type
according to the N/O ratio only.  They defined Type I PNe as those
PNe with N/O $\ge 0.8$. Following this criterion, we classify all
four PNe in our sample as non-Type I objects.

The Type I PNe were believed to have evolved from the most massive
progenitor stars with initial main sequence masses range 2.4-8.0
M$_{\bigodot}$ \citep{Quireza07} and consequently they should be
associated with the thin Galactic disk and have low velocity
dispersion. They also show a wide range of ionisation structures.
Using the adopted distance and measured radial velocity, we derived
a vertical Galactic height $z = 210$ pc and a peculiar
velocity\footnote{The difference between the observed radial
velocity corrected for local standard of rest (LSR) and the velocity
determined from the Galactic rotation curve.} $V_P = 19.4$
kms$^{-1}$ for Hen2-29. Previous measurements of the spatial ($z <
300$ pc, \citet{Gilmore83}) and kinematic characteristics ($V_P <
60$ kms$^{-1}$, \citet{Maciel92}) of this PNe show that it belongs
to the thin Galactic disk population.

\citet{Faundez87} developed the Peimbert classification scheme by
dividing the Peimbert Type II into Type IIa and Type IIb. This
division was proposed based essentially on the nitrogen abundances.
They show that Type IIa are of nitrogen abundances (log (N/H)+12
$\geq 8.0$) larger than that of Type IIb (log (N/H)+12 $\leq 8.0$).
Following the further re-analysis of Peimbert types suggested by
\citet{Quireza07}, we find both M3-4 and Hen2-37 are consistent with
being of Type IIa ($\log(\rm N/H)+12 \geq 8.0$ and N/O $\geq 0.25$,
but He/H $\leq 0.125$). Therefore, they should arise from an older
and less massive main sequence population (1.2-2.4 M$_{\bigodot}$)
than the Type I objects. Furthermore, the spatial and kinematic
properties of M3-4 (336 pc and 29.9 km s$^{-1}$) and Hen2-37(185 pc
and 7.3 km s$^{-1}$) show that they are thin Galactic disk members.
M 3-6 nebula is the only PN in the set that shows ($\log(\rm N/H)+12
< 8.0$). The object lies at $z = 144$ pc and has $V_P = 7.3$ km
s$^{-1}$, hence it can be classified as a Type IIb PN.

\section{Kinematical parameters and distances}

\begin{table*}
\centering \caption{Radial and expansion velocities and distances of
the sample}
 \label{Table7}
\scalebox{1.0}{
\begin{tabular}{lcccccccccc}
 \hline
  Object    & \multicolumn{2}{|c|}{$RV_{\rm hel}$ (km S$^{-1}$)}&
 \multicolumn{2}{|c|}{V$_{exp}$ (km S$^{-1}$)} &  \multicolumn{2}{|c|}{Distance (kpc)} & \multicolumn{2}{|c|}{Excitation class}\\
      &    Ours    &   literature     & [S II] & [N II]  &  (I) & (II) & (I) & (II)\\
\hline
M3-4          & 25.0$\pm$3.24   & 74$\pm$40 (1), 35.4$\pm$4.6(3)  & 18.7   & 17.5 & 3.82 &  4.27 & 6.2 & 7.8\\
M3-6       & 31.1$\pm$3.5 & 49.7$\pm$16.6 (1) & 22.2   & 28.3 & 3.05 &  3.05 & 4.6 & 4.6 \\
Hen2-29           & 19.8$\pm$2.9 &  25$\pm$6 (2)   & 33.8    & 29.1      & 4.02 &  4.32 & 8.0 & 10.5\\
Hen2-37        & -12.5$\pm$3.12  & 12$\pm$5 (2)    & 47.9     & 52.2 & 3.35 &  4.01 & 8.4 & 11.4 \\
 \hline
\end{tabular}}
\begin{minipage}[!t]{14cm}
{\tiny {\bf References:} (1) \citet{Schneider83}, (2) \citet{Meatheringham88}, (3) \citet{Kniazev12}\\
{\bf Distances:} scales; (I) \citet{Ali15a}; (II) \citet{Frew16}.\\
{\bf Excitation class:} (I) \citet{Meatheringham91}; (II)
\citet{Reid10}.}
\end{minipage}
\end{table*}

In order to detemine the evolutionary status of a PN, we need its
expansion velocity ($V_{exp}$) and its  distance. Here, we
determined $V_{exp}$ from the two emission lines [S\,II] and [N\,II]
which lie in the red channel of WiFeS instrument with the higher
spectral resolution ($R \sim 7000$). The FWHM of each line was
measured using the IRAF {\tt splot} task. The expansion velocity was
corrected for instrumental and thermal broadening following
\citep{Gieseking86}. The results were listed in Table \ref{Table7}.

Hen2-37 has the highest expansion velocity, and in this object
double  peak nebular emission lines are clearly seen. For Hen2-29 we
derive $V_{exp}$ -- greater than the standard value of 20 kms$^{-1}$
\citep{Weinberger89}.  \citet{Meatheringham88} report expansion
velocities of 23.6 kms$^{-1}$ and 30.7 kms$^{-1}$ for Hen2-29  and
Hen2-37, respectively. These are smaller than we determine, but this
is to be expected since the  \citet{Meatheringham88} results rely on
the [O\,III] emission line which has higher ionisation potential
than the lines we used, and is therefore produced closer to the
centre of the nebula - see the central panels Figures \ref{fig3} and
\ref{fig4}.

The systemic velocities $RV_{\rm sys}$ of the sample were measured
using the {\tt emsao} task  of the IRAF package. The $RV_{\rm sys}$
were derived from the Doppler-shift of [N~II] 6548\AA{}, $H\alpha$
6562\AA{}, [N\,II] 6583\AA{}, He I 6678\AA{}, [S\,II] 6716\AA{} and
[S\,II] 6730\AA{} emission lines. We select these lines because they
are locate in the high spectral resolution part of nebular spectra
($R \sim 7000$). To derive the heliocentric radial velocity $RV_{\rm
hel}$, we used the IRAF task {\tt RVCorr} to correct for the effect
of the Earth's motion around the Sun. The results were presented in
Table \ref{Table7} and compared with the work of
(\citet{Schneider83}, hereafter STPP83), \citet{Meatheringham88} and
\citet{Kniazev12}. In general the determined $RV_{\rm hel}$ of M3-4
nebula is relatively close to the recent estimation of
\citet{Kniazev12} but it is smaller than of STPP83. The value of
STPP83 was originally taken from the low dispersion spectra of
``Mayall-1964" given as a private communication with Perek and
Kohoutek (1967).  The $RV_{\rm hel}$ for M3-6 differs significantly
in accuracy with STPP83. The value of STPP83 was taken as a weighted
average of two unpublished measurements 57$\pm$11 (Minkowski 1957)
and 12$\pm$25 (Mayall 1964) as a private communication with Perek \&
Kohoutek (1967). This average value was weighted according to the
error of each measurement. For the object Hen2-29, our derived
$RV_{\rm hel}$ is consistent with that of MWF88 within the error
range. For Hen2-37, we determined $RV_{\rm hel}$ of -12.5$\pm$3.12
which is differs significantly than that of \citet{Meatheringham88}.
Our value was checked by measuring $RV_{\rm hel}$ = -16.0$\pm$5.0,
from the blue part of the nebular spectrum (which has lower
resolution, $R=3000$, than the red).  Three possible explanations
can be provided for this discrepancy: (1) The different spectrum
resolution particularly most of nebular lines appear of double peaks
due to the high expansion velocity of the nebula; (2) The radial
velocity determined here was measured from integrated spectrum over
the whole object; (3) The negative sign of the radial velocity
measure given by \citet{Meatheringham88} was simply missed.

None of the sample has a distance determined from either the
trigonometric, spectroscopic, cluster membership, or expansion
methods. Therefore, we must rely on the statistical approaches to
estimate distances, recognising the large errors that this entails.
We adopt here the average distance derived from the recent two
distance scales of \citet{Ali15a} and \citet{Frew16}, for each PN.
The \citet{Ali15a}  scale depends on  the mass-radius and radio
surface brightness temperature-radius empirical relationships, and
specifically on the nebular angular size and 5 GHz radio flux. The
\citet{Frew16} scale depends on the empirical relationship between
H$\alpha$ surface brightness and the radius of PN, using the nebular
angular size and the H$\alpha$ flux. The results of the two
approaches are given in Table \ref{Table6}.  Both methods give
roughly the same distances. The angular radii of the PNe are taken
from \citet{Frew16}, while radio fluxes at 5GHz were taken from
\citet{Cahn92} except for M3-4. We find this object has radio fluxes
which differ between the different references. We adopted here the
average value from \citet{Milne75}, \citet{Milne79},
\citet{Zijlstra89} and \citet{Cahn92}.

\section{The central star of M3-6}
\citet{Tylenda93}, \citet{Marcolino03} and \citet{Weidmann11}
classified the CS of the M3-6 nebula as being of the WELS type.
\citet{Miszalski09} reported that many of WELS are probably
misclassified close binaries. Further \citet{Miszalski11} and
\citet{Corradi11} observed many of WELS emission lines in the
spectra of stars known to be close binary systems, and explained
that these lines were originated from the irradiated zone on the
side of the companion facing the primary.

\begin{figure*}
\includegraphics[scale=1.4]{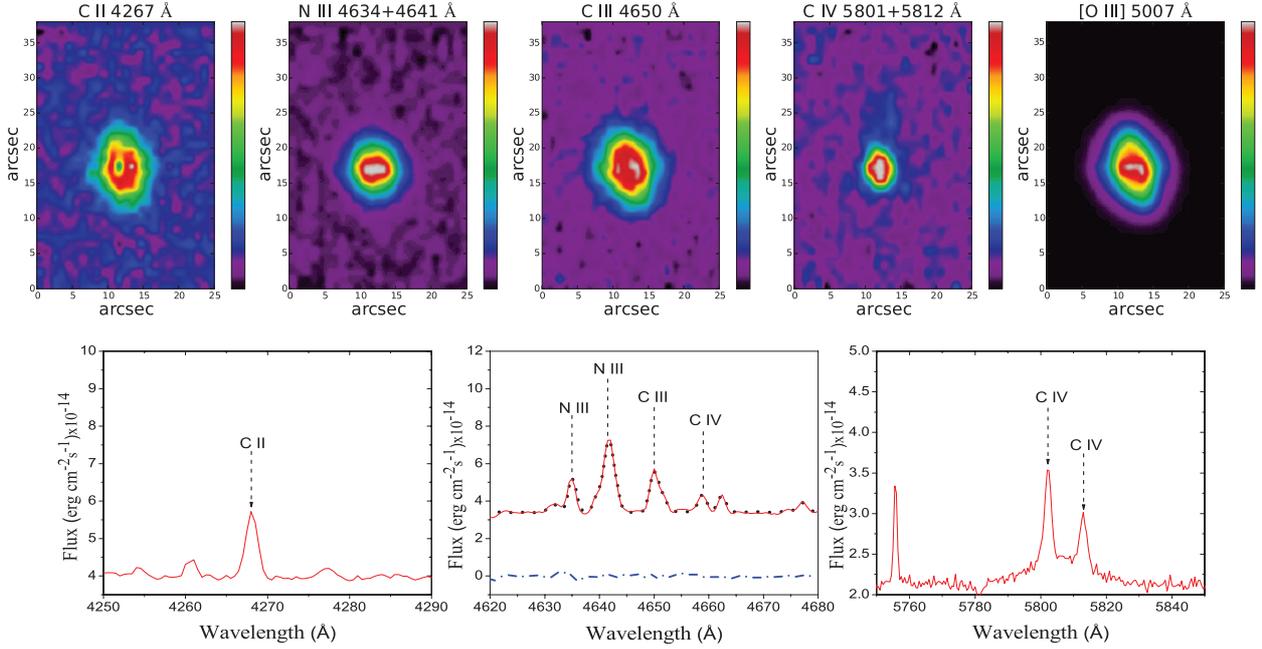}

  \caption{Emission-line maps of M 3-6 nebula in the recombination lines: CII at 4267\AA{};
  NIII at 4634\AA{}+4641\AA{}; CIII at 4650\AA{}; CIV at 5801\AA{}+5812\AA{}
  (sorted from left to right).  All emission-line maps reveal that these recombination lines
  are spatially extended, and therefore they originate from the nebulae, with the exception
  of CIV at 5801\AA{}+5812\AA{} which is probably emitted from the central nebular star. In the right upper
  panel, we present an emission-line map of M 3-6 in the collisional excitation line of [OIII] at
  5007\AA{} for comparison. The lower panels show the characteristic emission lines (in red color)
  from the M 3-6 spectrum which are generally used to classify nebular central star as a WELS type.
In the middle lower panel the black dot line
  shows the Gaussian fit of the observed N III, C III, and C IV recombination nebular lines, while the blue dash-dot line shows
  the residual between the observed and fitted fluxes.}
  \label{fig5}
\end{figure*}

\citet{Basurah16} claimed that, for many objects, the WELS class of
the PNe central star may be spurious. From WiFeS data of NGC 5979,
M4-2 and My60, they showed that the characteristic CS recombination
lines of WELS type are of nebular origin. Specifically, the lines
used to provide the WELS classification are CII at 4267\AA{}, NIII
at 4634\AA{}, 4641\AA{}, CIII at 4650\AA{}, CIV at 5801\AA{} and
5812\AA{}. The CS classification and the WELs class in general were
discussed in detail by \citet{Basurah16}. Here, we have found a
further example that increases the doubt regarding the reliability
of the WELS classification. In Figure \ref{fig5} we show emission
line maps of M3-6 in four CS emission lines, CII, CIII, CIV, and
NIII used to define the WELS class. It is apparent that the emission
of these lines are spatial distributed over a large area of the
nebula, and hence are clearly of nebular rather than of CS origin.
Only in CIV at 5801\AA{} and 5812\AA{} is the emission limited to
the central nebular region and consequently probably originates on
the CS.

\begin{figure*}
  \includegraphics[scale=0.6]{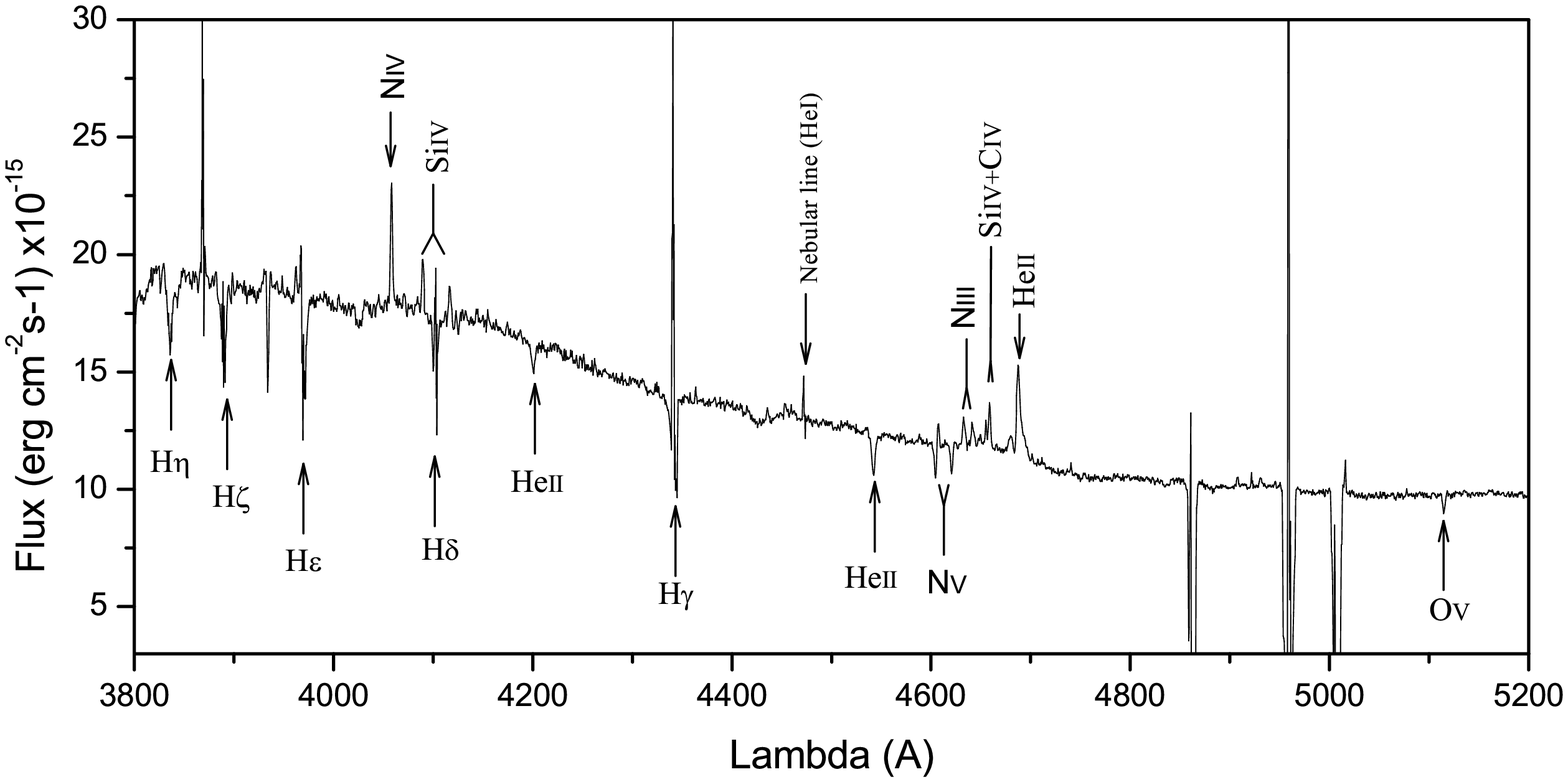}
   \includegraphics[scale=0.6]{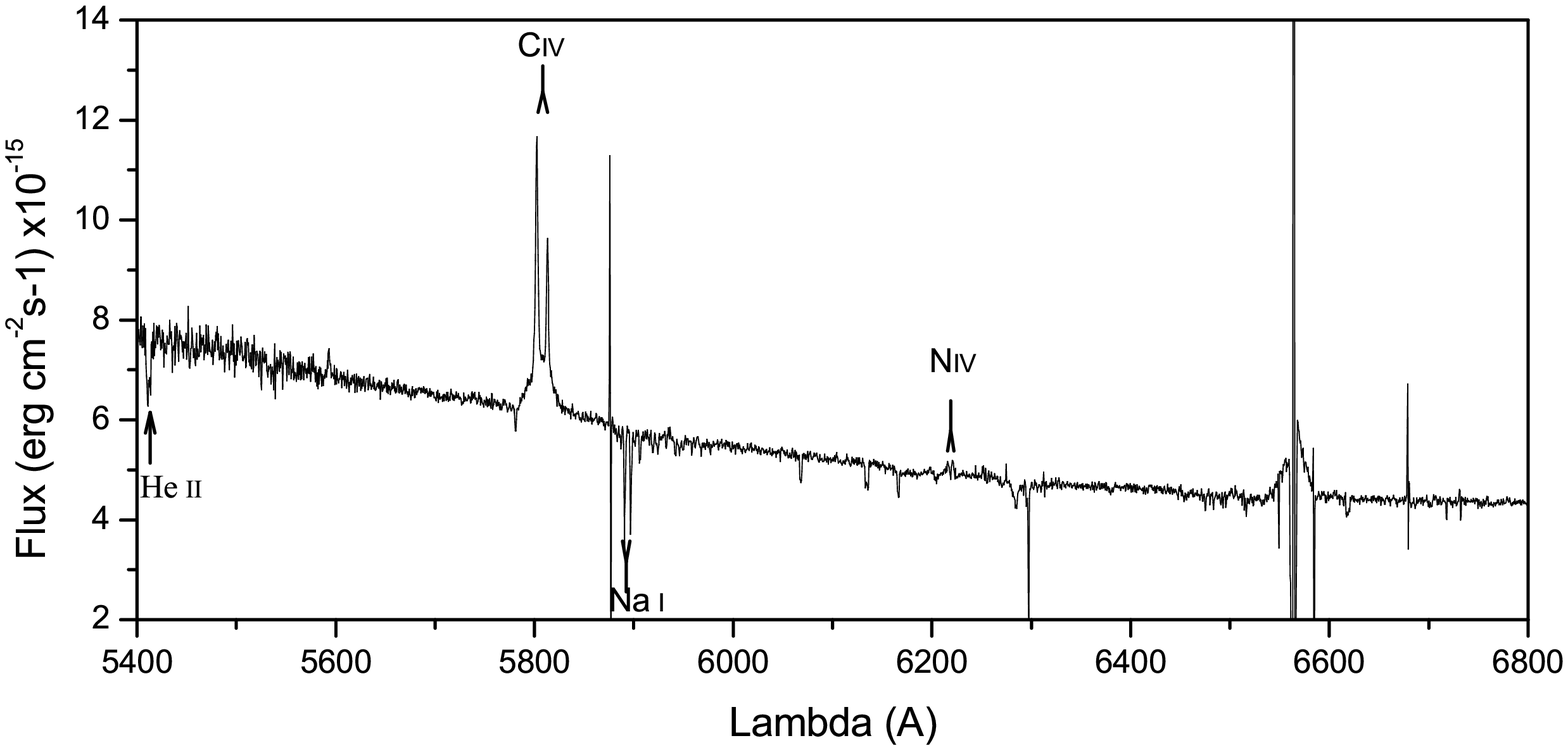}
  \caption{The nebular subtracted spectrum of the CS in M3-6.
  The blue spectrum (upper panel) shows clearly the presence of Balmer and He
  II lines in absorbtion, with the exception of He II $\lambda$ 4686 which is seen in
  emission. The red spectrum (lower panel) is characterised by the presence of
  C IV  doublet lines at $\lambda \lambda 5801, 5811$, N IV at
  $\lambda \lambda6212, 6220$, and the interstellar Na I D lines at $\lambda
  \lambda 5890,5896$ seen in absorption. More discussion on this CS spectrum
  and the stellar classification derived from it is given in the text.}\label{fig6}
\end{figure*}

In fact, we can provide a revised classification for this central
star based upon our own data. In Figure \ref{fig6} we show our
complete extracted  spectrum of the central star of M3-6. To obtain
this, we carefully removed the nebular emission determined from a
zone around the CS from the spaxels defining the continuum image of
the CS in the image cube. Many Balmer and He II lines clearly
visible in absorption in the blue spectrum (upper panel), except He
II at $\lambda$4686 appears in emission. Further, many other
emission lines are present such as N IV $\lambda$4058, Si IV
$\lambda$4089, $\lambda$4116, $\lambda$4654, C IV $\lambda$4658, and
N III $\lambda\lambda$4634, 4640.
 The N V doublet $\lambda\lambda 4604,4620$ and the O V  line at
 $\lambda$5114  both appear in absorption. In the red spectrum (lower panel),
  the  C IV doublet $\lambda \lambda 5801,5811$ and the
  N IV doublet $\lambda\lambda6212,6220$ are clearly seen in emission,
   while He II $\lambda$5412 and the interstellar  lines of Na I D
   $\lambda\lambda 5890,5896$ are visible in absorption.

  Following the CS classification scheme of \citet{Mendez91} and
  improvements given to this scheme by \citet{Weidmann15}, we classify
  the CS of M3-6 nebula as a H-rich star of spectral type O3 I(f*).
  The relative weakness of the N V  $\lambda\lambda 4604,4620$ doublet
  compared to the He II $\lambda4541$ line is an indication of a spectral
  type O3, and the presence of N III the $\lambda\lambda4634,4640$
  doublet along with the He II $\lambda4686$ line are features of
Of(H) type. Further, the stronger emission of N IV  $\lambda4058$
relative to N III $\lambda$4640 provides the * qualifier which is a
unique property stars of the O3 type \citep{Walborn20}. In general,
the Si IV emission is also present in Of* spectra. From Figure
\ref{fig6}, it appears that the C IV recombination lines at $\lambda
\lambda5801,5811$ are of CS origin -- as is also revealed in the
emission line map of the nebula constructed in these two lines and
shown in Figure \ref{fig5}.

\section{Conclusions}
In this paper we presented the first integral field spectroscopy of
the southern planetary nebulae M3-4, M3-6, Hen2-29 and Hen2-37 in
the optical range 3400-7000 \AA{}. We demonstrated the utility of
these observations in both providing narrow-band data to probe the
morphological and excitation structure of the nebulae as well as in
deriving their dynamical nature, their optical thickness and their
chemical enrichment characteristics.

The four PNe have different optical thickness, where M3-6 is
optically thin, Hen2-37 is optically thick while M3-4 and Hen2-29
are partially optically thick. From the strength of He II line, we
derived excitation class 6.2-7.8, 4.6, 8.0-10.5, and 8.4-11.4 for
M3-4, M3-6, Hen2-29 and Hen2-37, respectively. From the chemical
analysis of the sample based upon integral spectroscopy, we provided
Peimbert types I, IIa, IIa, IIb for Hen2-29, M3-4, Hen2-37 and M3-6,
respectively, and noted that the long-slit spectroscopic data can
provide surprisingly good results even though only a sub-region of
the PN is analysed by this technique.

In the case of M3-6, we find that the majority of the recombination
lines used in literature to classify the CS as a weak emission-line
star are in fact of nebular origin. Instead we classify the central
star as  H-rich and of spectral type O3 I(f*). This result extends
to five (M3-6 and NGC 3211, NGC 5979, My 60, M 4-2 that mentioned in
\citet{Basurah16}) the number of cases of mis-classification of WELs
stars discovered using
 integral field spectroscopy and served to increase doubts regarding
 the reliability of the WELS classification in general. In this we
 support the conclusions of \citet{Weidmann15} who observed 19 objects with
 the WELS classification amongst the total of 72 so classified. They
 concluded that ""the denomination WELS should not be taken as a spectral
 type, because, as a WELS is based on low-resolution spectra, it cannot
 provide enough information about the photospheric H abundance".

\bibliographystyle{mn2e_new}
\bibliography{PaperII_PNe}

\begin{thebibliography}{}
\makeatletter
\relax
\def\mn@urlcharsother{\let\do\@makeother \do\$\do\&\do\#\do\^\do\_\do\%\do\~}
\def\mn@doi{\begingroup\mn@urlcharsother \@ifnextchar [ {\mn@doi@}
  {\mn@doi@[]}}
\def\mn@doi@[#1]#2{\def\@tempa{#1}\ifx\@tempa\@empty \href
  {http://dx.doi.org/#2} {doi:#2}\else \href {http://dx.doi.org/#2} {#1}\fi
  \endgroup}
\def\mn@eprint#1#2{\mn@eprint@#1:#2::\@nil}
\def\mn@eprint@arXiv#1{\href {http://arxiv.org/abs/#1} {{\tt arXiv:#1}}}
\def\mn@eprint@dblp#1{\href {http://dblp.uni-trier.de/rec/bibtex/#1.xml}
  {dblp:#1}}
\def\mn@eprint@#1:#2:#3:#4\@nil{\def\@tempa {#1}\def\@tempb {#2}\def\@tempc
  {#3}\ifx \@tempc \@empty \let \@tempc \@tempb \let \@tempb \@tempa \fi \ifx
  \@tempb \@empty \def\@tempb {arXiv}\fi \@ifundefined
  {mn@eprint@\@tempb}{\@tempb:\@tempc}{\expandafter \expandafter \csname
  mn@eprint@\@tempb\endcsname \expandafter{\@tempc}}}

\bibitem[\protect\citeauthoryear{{Acker}, {Raytchev}, {Stenholm}  \&
  {Tylenda}}{{Acker} et~al.}{1991}]{Acker91}
{Acker} A.,  {Raytchev} B.,  {Stenholm} B.,   {Tylenda} R.,  1991, \aaps, \href
  {http://adsabs.harvard.edu/abs/1991A%26AS...90...89A} {90, 89}

\bibitem[\protect\citeauthoryear{{Akras} \& {Gon{\c c}alves}}{{Akras} \&
  {Gon{\c c}alves}}{2016}]{Akras16a}
{Akras} S.,  {Gon{\c c}alves} D.~R.,  2016, \mn@doi [\mnras]
  {10.1093/mnras/stv2139}, \href
  {http://adsabs.harvard.edu/abs/2016MNRAS.455..930A} {455, 930}

\bibitem[\protect\citeauthoryear{{Akras}, {Clyne}, {Boumis}, {Monteiro},
  {Gon{\c c}alves}, {Redman}  \& {Williams}}{{Akras} et~al.}{2016}]{Akras16b}
{Akras} S.,  {Clyne} N.,  {Boumis} P.,  {Monteiro} H.,  {Gon{\c c}alves} D.~R.,
   {Redman} M.~P.,   {Williams} S.,  2016, \mn@doi [\mnras]
  {10.1093/mnras/stw038}, \href
  {http://adsabs.harvard.edu/abs/2016MNRAS.457.3409A} {457, 3409}

\bibitem[\protect\citeauthoryear{{Ali}, {Ismail}  \& {Alsolami}}{{Ali}
  et~al.}{2015a}]{Ali15a}
{Ali} A.,  {Ismail} H.~A.,   {Alsolami} Z.,  2015a, \mn@doi [\apss]
  {10.1007/s10509-015-2293-8}, \href
  {http://adsabs.harvard.edu/abs/2015Ap%26SS.357...21A} {357, 21}

\bibitem[\protect\citeauthoryear{{Ali}, {Amer}, {Dopita}, {Vogt}  \&
  {Basurah}}{{Ali} et~al.}{2015b}]{Ali15b}
{Ali} A.,  {Amer} M.~A.,  {Dopita} M.~A.,  {Vogt} F.~P.~A.,   {Basurah} H.~M.,
  2015b, \mn@doi [\aap] {10.1051/0004-6361/201526223}, \href
  {http://adsabs.harvard.edu/abs/2015A%26A...583A..83A} {583, A83}

\bibitem[\protect\citeauthoryear{{Aller}, {Miranda}, {Olgu{\'{\i}}n},
  {V{\'a}zquez}, {Guill{\'e}n}, {Oreiro}, {Ulla}  \& {Solano}}{{Aller}
  et~al.}{2015}]{Aller15}
{Aller} A.,  {Miranda} L.~F.,  {Olgu{\'{\i}}n} L.,  {V{\'a}zquez} R.,
  {Guill{\'e}n} P.~F.,  {Oreiro} R.,  {Ulla} A.,   {Solano} E.,  2015, \mn@doi
  [\mnras] {10.1093/mnras/stu2106}, \href
  {http://adsabs.harvard.edu/abs/2015MNRAS.446..317A} {446, 317}

\bibitem[\protect\citeauthoryear{{Balick}, {Preston}  \& {Icke}}{{Balick}
  et~al.}{1987}]{Balick87}
{Balick} B.,  {Preston} H.~L.,   {Icke} V.,  1987, \mn@doi [\aj]
  {10.1086/114595}, \href {http://adsabs.harvard.edu/abs/1987AJ.....94.1641B}
  {94, 1641}

\bibitem[\protect\citeauthoryear{{Balick}, {Rugers}, {Terzian}  \&
  {Chengalur}}{{Balick} et~al.}{1993}]{Balick93}
{Balick} B.,  {Rugers} M.,  {Terzian} Y.,   {Chengalur} J.~N.,  1993, \mn@doi
  [\apj] {10.1086/172881}, \href
  {http://adsabs.harvard.edu/abs/1993ApJ...411..778B} {411, 778}

\bibitem[\protect\citeauthoryear{{Balick}, {Perinotto}, {Maccioni}, {Terzian}
  \& {Hajian}}{{Balick} et~al.}{1994}]{Balick94}
{Balick} B.,  {Perinotto} M.,  {Maccioni} A.,  {Terzian} Y.,   {Hajian} A.,
  1994, \mn@doi [\apj] {10.1086/173932}, \href
  {http://adsabs.harvard.edu/abs/1994ApJ...424..800B} {424, 800}

\bibitem[\protect\citeauthoryear{{Basurah}, {Ali}, {Dopita}, {Alsulami}, {Amer}
   \& {Alruhaili}}{{Basurah} et~al.}{2016}]{Basurah16}
{Basurah} H.~M.,  {Ali} A.,  {Dopita} M.~A.,  {Alsulami} R.,  {Amer} M.~A.,
  {Alruhaili} A.,  2016, \mn@doi [\mnras] {10.1093/mnras/stw468}, \href
  {http://adsabs.harvard.edu/abs/2016MNRAS.tmp..244B} {}

\bibitem[\protect\citeauthoryear{{Cahn}, {Kaler}  \& {Stanghellini}}{{Cahn}
  et~al.}{1992}]{Cahn92}
{Cahn} J.~H.,  {Kaler} J.~B.,   {Stanghellini} L.,  1992, \aaps, \href
  {http://adsabs.harvard.edu/abs/1992A%26AS...94..399C} {94, 399}

\bibitem[\protect\citeauthoryear{{Chiappini} \& {Maciel}}{{Chiappini} \&
  {Maciel}}{1994}]{Chiappini94}
{Chiappini} C.,  {Maciel} W.~J.,  1994, \aap, \href
  {http://adsabs.harvard.edu/abs/1994A%26A...288..921C} {288, 921}

\bibitem[\protect\citeauthoryear{{Childress}, {Vogt}, {Nielsen}  \&
  {Sharp}}{{Childress} et~al.}{2014}]{Childress14}
{Childress} M.~J.,  {Vogt} F.~P.~A.,  {Nielsen} J.,   {Sharp} R.~G.,  2014,
  \mn@doi [\apss] {10.1007/s10509-013-1682-0}, \href
  {http://adsabs.harvard.edu/abs/2014Ap%26SS.349..617C} {349, 617}

\bibitem[\protect\citeauthoryear{{Corradi}, {Aznar}  \& {Mampaso}}{{Corradi}
  et~al.}{1998}]{Corradi98}
{Corradi} R.~L.~M.,  {Aznar} R.,   {Mampaso} A.,  1998, \mn@doi [\mnras]
  {10.1046/j.1365-8711.1998.01532.x}, \href
  {http://adsabs.harvard.edu/abs/1998MNRAS.297..617C} {297, 617}

\bibitem[\protect\citeauthoryear{{Corradi} et~al.,}{{Corradi}
  et~al.}{2011}]{Corradi11}
{Corradi} R.~L.~M.,  et~al., 2011, \mn@doi [\mnras]
  {10.1111/j.1365-2966.2010.17523.x}, \href
  {http://adsabs.harvard.edu/abs/2011MNRAS.410.1349C} {410, 1349}

\bibitem[\protect\citeauthoryear{{Danehkar}}{{Danehkar}}{2015}]{Danehkar15x}
{Danehkar} A.,  2015, \mn@doi [\apj] {10.1088/0004-637X/815/1/35}, \href
  {http://adsabs.harvard.edu/abs/2015ApJ...815...35D} {815, 35}

\bibitem[\protect\citeauthoryear{{Danehkar} \& {Parker}}{{Danehkar} \&
  {Parker}}{2015}]{Danehkar15}
{Danehkar} A.,  {Parker} Q.~A.,  2015, \mn@doi [\mnras]
  {10.1093/mnrasl/slv022}, \href
  {http://adsabs.harvard.edu/abs/2015MNRAS.449L..56D} {449, L56}

\bibitem[\protect\citeauthoryear{{De Freitas Pacheco}, {Maciel}  \&
  {Costa}}{{De Freitas Pacheco} et~al.}{1992}]{de-Freitas92}
{De Freitas Pacheco} J.~A.,  {Maciel} W.~J.,   {Costa} R.~D.~D.,  1992, \aap,
  \href {http://adsabs.harvard.edu/abs/1992A%26A...261..579D} {261, 579}

\bibitem[\protect\citeauthoryear{{Delgado-Inglada}, {Morisset}  \&
  {Stasi{\'n}ska}}{{Delgado-Inglada} et~al.}{2014}]{Delgado14}
{Delgado-Inglada} G.,  {Morisset} C.,   {Stasi{\'n}ska} G.,  2014, in Revista
  Mexicana de Astronomia y Astrofisica Conference Series. pp 17--17

\bibitem[\protect\citeauthoryear{{Dopita}}{{Dopita}}{1997}]{Dopita97}
{Dopita} M.~A.,  1997, \mn@doi [\apjl] {10.1086/310804}, \href
  {http://adsabs.harvard.edu/abs/1997ApJ...485L..41D} {485, L41}

\bibitem[\protect\citeauthoryear{{Dopita}, {Hart}, {McGregor}, {Oates},
  {Bloxham}  \& {Jones}}{{Dopita} et~al.}{2007}]{Dopita07}
{Dopita} M.,  {Hart} J.,  {McGregor} P.,  {Oates} P.,  {Bloxham} G.,   {Jones}
  D.,  2007, \mn@doi [\apss] {10.1007/s10509-007-9510-z}, \href
  {http://adsabs.harvard.edu/abs/2007Ap%26SS.310..255D} {310, 255}

\bibitem[\protect\citeauthoryear{{Dopita} et~al.,}{{Dopita}
  et~al.}{2010}]{Dopita10}
{Dopita} M.,  et~al., 2010, \mn@doi [\apss] {10.1007/s10509-010-0335-9}, \href
  {http://adsabs.harvard.edu/abs/2010Ap%26SS.327..245D} {327, 245}

\bibitem[\protect\citeauthoryear{{Faundez-Abans} \& {Maciel}}{{Faundez-Abans}
  \& {Maciel}}{1987}]{Faundez87}
{Faundez-Abans} M.,  {Maciel} W.~J.,  1987, \aap, \href
  {http://adsabs.harvard.edu/abs/1987A%26A...183..324F} {183, 324}

\bibitem[\protect\citeauthoryear{{Frew}, {Boji{\v c}i{\'c}}  \&
  {Parker}}{{Frew} et~al.}{2013}]{Frew13}
{Frew} D.~J.,  {Boji{\v c}i{\'c}} I.~S.,   {Parker} Q.~A.,  2013, \mn@doi
  [\mnras] {10.1093/mnras/sts393}, \href
  {http://adsabs.harvard.edu/abs/2013MNRAS.431....2F} {431, 2}

\bibitem[\protect\citeauthoryear{{Frew}, {Parker}  \& {Boji{\v
  c}i{\'c}}}{{Frew} et~al.}{2016}]{Frew16}
{Frew} D.~J.,  {Parker} Q.~A.,   {Boji{\v c}i{\'c}} I.~S.,  2016, \mn@doi
  [\mnras] {10.1093/mnras/stv1516}, \href
  {http://adsabs.harvard.edu/abs/2016MNRAS.455.1459F} {455, 1459}

\bibitem[\protect\citeauthoryear{{Garc{\'{\i}}a-Rojas}, {Corradi}, {Monteiro},
  {Jones}, {Rodr{\'{\i}}guez-Gil}  \& {Cabrera-Lavers}}{{Garc{\'{\i}}a-Rojas}
  et~al.}{2016}]{Garcia16}
{Garc{\'{\i}}a-Rojas} J.,  {Corradi} R.~L.~M.,  {Monteiro} H.,  {Jones} D.,
  {Rodr{\'{\i}}guez-Gil} P.,   {Cabrera-Lavers} A.,  2016, e-print
  ArXiv:1606.02830, \href {http://adsabs.harvard.edu/abs/2016arXiv160602830G}
  {}

\bibitem[\protect\citeauthoryear{{Gieseking}, {Hippelein}  \&
  {Weinberger}}{{Gieseking} et~al.}{1986}]{Gieseking86}
{Gieseking} F.,  {Hippelein} H.,   {Weinberger} R.,  1986, \aap, \href
  {http://adsabs.harvard.edu/abs/1986A%26A...156..101G} {156, 101}

\bibitem[\protect\citeauthoryear{{Gilmore} \& {Reid}}{{Gilmore} \&
  {Reid}}{1983}]{Gilmore83}
{Gilmore} G.,  {Reid} N.,  1983, \mnras, \href
  {http://adsabs.harvard.edu/abs/1983MNRAS.202.1025G} {202, 1025}

\bibitem[\protect\citeauthoryear{{Girard}, {K{\"o}ppen}  \& {Acker}}{{Girard}
  et~al.}{2007}]{Girard07}
{Girard} P.,  {K{\"o}ppen} J.,   {Acker} A.,  2007, \mn@doi [\aap]
  {10.1051/0004-6361:20052807}, \href
  {http://adsabs.harvard.edu/abs/2007A%26A...463..265G} {463, 265}

\bibitem[\protect\citeauthoryear{{G{\'o}rny}, {Schwarz}, {Corradi}  \& {Van
  Winckel}}{{G{\'o}rny} et~al.}{1999}]{Gorny99}
{G{\'o}rny} S.~K.,  {Schwarz} H.~E.,  {Corradi} R.~L.~M.,   {Van Winckel} H.,
  1999, \mn@doi [\aaps] {10.1051/aas:1999205}, \href
  {http://adsabs.harvard.edu/abs/1999A%26AS..136..145G} {136, 145}

\bibitem[\protect\citeauthoryear{{Guerrero}, {Toal{\'a}}, {Medina},
  {Luridiana}, {Miranda}, {Riera}  \& {Vel{\'a}zquez}}{{Guerrero}
  et~al.}{2013}]{Guerrero13}
{Guerrero} M.~A.,  {Toal{\'a}} J.~A.,  {Medina} J.~J.,  {Luridiana} V.,
  {Miranda} L.~F.,  {Riera} A.,   {Vel{\'a}zquez} P.~F.,  2013, \mn@doi [\aap]
  {10.1051/0004-6361/201321786}, \href
  {http://adsabs.harvard.edu/abs/2013A%26A...557A.121G} {557, A121}

\bibitem[\protect\citeauthoryear{{Hajian} et~al.,}{{Hajian}
  et~al.}{2007}]{Hajian07}
{Hajian} A.~R.,  et~al., 2007, \mn@doi [\apjs] {10.1086/511767}, \href
  {http://adsabs.harvard.edu/abs/2007ApJS..169..289H} {169, 289}

\bibitem[\protect\citeauthoryear{{Howarth}}{{Howarth}}{1983}]{Howarth83}
{Howarth} I.~D.,  1983, \mnras, \href
  {http://adsabs.harvard.edu/abs/1983MNRAS.203..301H} {203, 301}

\bibitem[\protect\citeauthoryear{{Kaler} \& {Jacoby}}{{Kaler} \&
  {Jacoby}}{1989}]{Kaler89}
{Kaler} J.~B.,  {Jacoby} G.~H.,  1989, \mn@doi [\apj] {10.1086/167957}, \href
  {http://adsabs.harvard.edu/abs/1989ApJ...345..871K} {345, 871}

\bibitem[\protect\citeauthoryear{{Kingsburgh} \& {Barlow}}{{Kingsburgh} \&
  {Barlow}}{1994}]{Kingsburgh94}
{Kingsburgh} R.~L.,  {Barlow} M.~J.,  1994, \mnras, \href
  {http://adsabs.harvard.edu/abs/1994MNRAS.271..257K} {271, 257}

\bibitem[\protect\citeauthoryear{{Kniazev}}{{Kniazev}}{2012}]{Kniazev12}
{Kniazev} A.~Y.,  2012, \mn@doi [Astronomy Letters]
  {10.1134/S1063773712100040}, \href
  {http://adsabs.harvard.edu/abs/2012AstL...38..707K} {38, 707}

\bibitem[\protect\citeauthoryear{{Maciel} \& {Dutra}}{{Maciel} \&
  {Dutra}}{1992}]{Maciel92}
{Maciel} W.~J.,  {Dutra} C.~M.,  1992, \aap, \href
  {http://adsabs.harvard.edu/abs/1992A%26A...262..271M} {262, 271}

\bibitem[\protect\citeauthoryear{{Maciel} \& {Quireza}}{{Maciel} \&
  {Quireza}}{1999}]{Maciel99}
{Maciel} W.~J.,  {Quireza} C.,  1999, \aap, \href
  {http://adsabs.harvard.edu/abs/1999A%26A...345..629M} {345, 629}

\bibitem[\protect\citeauthoryear{{Manchado}, {Guerrero}, {Stanghellini}  \&
  {Serra-Ricart}}{{Manchado} et~al.}{1996}]{Manchado96}
{Manchado} A.,  {Guerrero} M.~A.,  {Stanghellini} L.,   {Serra-Ricart} M.,
  1996, {The IAC morphological catalog of northern Galactic planetary nebulae}

\bibitem[\protect\citeauthoryear{{Marcolino} \& {de Ara{\'u}jo}}{{Marcolino} \&
  {de Ara{\'u}jo}}{2003}]{Marcolino03}
{Marcolino} W.~L.~F.,  {de Ara{\'u}jo} F.~X.,  2003, \mn@doi [\aj]
  {10.1086/375908}, \href {http://adsabs.harvard.edu/abs/2003AJ....126..887M}
  {126, 887}

\bibitem[\protect\citeauthoryear{{Martins} \& {Viegas}}{{Martins} \&
  {Viegas}}{2000}]{Martins00}
{Martins} L.~P.,  {Viegas} S.~M.~M.,  2000, \aap, \href
  {http://adsabs.harvard.edu/abs/2000A%26A...361.1121M} {361, 1121}

\bibitem[\protect\citeauthoryear{{Meatheringham} \& {Dopita}}{{Meatheringham}
  \& {Dopita}}{1991}]{Meatheringham91}
{Meatheringham} S.~J.,  {Dopita} M.~A.,  1991, \mn@doi [\apjs]
  {10.1086/191536}, \href {http://adsabs.harvard.edu/abs/1991ApJS...75..407M}
  {75, 407}

\bibitem[\protect\citeauthoryear{{Meatheringham}, {Wood}  \&
  {Faulkner}}{{Meatheringham} et~al.}{1988}]{Meatheringham88}
{Meatheringham} S.~J.,  {Wood} P.~R.,   {Faulkner} D.~J.,  1988, \mn@doi [\apj]
  {10.1086/166882}, \href {http://adsabs.harvard.edu/abs/1988ApJ...334..862M}
  {334, 862}

\bibitem[\protect\citeauthoryear{{Mendez}}{{Mendez}}{1991}]{Mendez91}
{Mendez} R.~H.,  1991, in {Crivellari} L.,  {Hubeny} I.,   {Hummer} D.~G.,
  eds,  NATO Advanced Science Institutes (ASI) Series C Vol. 341, NATO Advanced
  Science Institutes (ASI) Series C. p.~331

\bibitem[\protect\citeauthoryear{{Mendoza}}{{Mendoza}}{1983}]{Mendoza83}
{Mendoza} C.,  1983, in {Flower} D.~R.,  ed.,  IAU Symposium Vol. 103,
  Planetary Nebulae. pp 143--172

\bibitem[\protect\citeauthoryear{{Milingo}, {Henry}  \& {Kwitter}}{{Milingo}
  et~al.}{2002}]{Milingo02}
{Milingo} J.~B.,  {Henry} R.~B.~C.,   {Kwitter} K.~B.,  2002, \mn@doi [\apjs]
  {10.1086/324292}, \href {http://adsabs.harvard.edu/abs/2002ApJS..138..285M}
  {138, 285}

\bibitem[\protect\citeauthoryear{{Milne}}{{Milne}}{1979}]{Milne79}
{Milne} D.~K.,  1979, \aaps, \href
  {http://cdsads.u-strasbg.fr/abs/1979A%26AS...36..227M} {36, 227}

\bibitem[\protect\citeauthoryear{{Milne} \& {Aller}}{{Milne} \&
  {Aller}}{1975}]{Milne75}
{Milne} D.~K.,  {Aller} L.~H.,  1975, \aap, \href
  {http://cdsads.u-strasbg.fr/abs/1975A%26A....38..183M} {38, 183}

\bibitem[\protect\citeauthoryear{{Miranda}, {Ramos-Larios}  \&
  {Guerrero}}{{Miranda} et~al.}{2010}]{Miranda10}
{Miranda} L.~F.,  {Ramos-Larios} G.,   {Guerrero} M.~A.,  2010, \mn@doi [\pasa]
  {10.1071/AS09028}, \href {http://adsabs.harvard.edu/abs/2010PASA...27..180M}
  {27, 180}

\bibitem[\protect\citeauthoryear{{Miszalski}}{{Miszalski}}{2009}]{Miszalski09}
{Miszalski} B.,  2009, PhD thesis, Department of Physics, Macquarie University,
  NSW 2109, Australia

\bibitem[\protect\citeauthoryear{{Miszalski}, {Corradi}, {Boffin}, {Jones},
  {Sabin}, {Santander-Garc{\'{\i}}a}, {Rodr{\'{\i}}guez-Gil}  \&
  {Rubio-D{\'{\i}}ez}}{{Miszalski} et~al.}{2011}]{Miszalski11}
{Miszalski} B.,  {Corradi} R.~L.~M.,  {Boffin} H.~M.~J.,  {Jones} D.,  {Sabin}
  L.,  {Santander-Garc{\'{\i}}a} M.,  {Rodr{\'{\i}}guez-Gil} P.,
  {Rubio-D{\'{\i}}ez} M.~M.,  2011, \mn@doi [\mnras]
  {10.1111/j.1365-2966.2011.18212.x}, \href
  {http://adsabs.harvard.edu/abs/2011MNRAS.413.1264M} {413, 1264}

\bibitem[\protect\citeauthoryear{{Monreal-Ibero}, {Roth}, {Sch{\"o}nberner},
  {Steffen}  \& {B{\"o}hm}}{{Monreal-Ibero} et~al.}{2005}]{Monreal-Ibero05}
{Monreal-Ibero} A.,  {Roth} M.~M.,  {Sch{\"o}nberner} D.,  {Steffen} M.,
  {B{\"o}hm} P.,  2005, \mn@doi [\apjl] {10.1086/432664}, \href
  {http://adsabs.harvard.edu/abs/2005ApJ...628L.139M} {628, L139}

\bibitem[\protect\citeauthoryear{{Peimbert} \& {Torres-Peimbert}}{{Peimbert} \&
  {Torres-Peimbert}}{1983}]{Peimbert83}
{Peimbert} M.,  {Torres-Peimbert} S.,  1983, in {Flower} D.~R.,  ed.,  IAU
  Symposium Vol. 103, Planetary Nebulae. pp 233--241

\bibitem[\protect\citeauthoryear{{Perinotto}, {Morbidelli}  \&
  {Scatarzi}}{{Perinotto} et~al.}{2004}]{Perinotto04}
{Perinotto} M.,  {Morbidelli} L.,   {Scatarzi} A.,  2004, \mn@doi [\mnras]
  {10.1111/j.1365-2966.2004.07470.x}, \href
  {http://adsabs.harvard.edu/abs/2004MNRAS.349..793P} {349, 793}

\bibitem[\protect\citeauthoryear{{Quireza}, {Rocha-Pinto}  \&
  {Maciel}}{{Quireza} et~al.}{2007}]{Quireza07}
{Quireza} C.,  {Rocha-Pinto} H.~J.,   {Maciel} W.~J.,  2007, \mn@doi [\aap]
  {10.1051/0004-6361:20078087}, \href
  {http://adsabs.harvard.edu/abs/2007A%26A...475..217Q} {475, 217}

\bibitem[\protect\citeauthoryear{{Reid} \& {Parker}}{{Reid} \&
  {Parker}}{2010}]{Reid10}
{Reid} W.~A.,  {Parker} Q.~A.,  2010, \mn@doi [\pasa] {10.1071/AS09055}, \href
  {http://adsabs.harvard.edu/abs/2010PASA...27..187R} {27, 187}

\bibitem[\protect\citeauthoryear{{Schneider} \& {Terzian}}{{Schneider} \&
  {Terzian}}{1983}]{Schneider83}
{Schneider} S.~E.,  {Terzian} Y.,  1983, \mn@doi [\apjl] {10.1086/184151},
  \href {http://adsabs.harvard.edu/abs/1983ApJ...274L..61S} {274, L61}

\bibitem[\protect\citeauthoryear{{Steffen}, {L{\'o}pez}  \& {Lim}}{{Steffen}
  et~al.}{2002}]{Steffen02}
{Steffen} W.,  {L{\'o}pez} J.~A.,   {Lim} A.~J.,  2002, in {Henney} W.~J.,
  {Steffen} W.,  {Binette} L.,   {Raga} A.,  eds,  Revista Mexicana de
  Astronomia y Astrofisica Conference Series Vol. 13, Revista Mexicana de
  Astronomia y Astrofisica Conference Series. pp 150--154

\bibitem[\protect\citeauthoryear{{Storey} \& {Zeippen}}{{Storey} \&
  {Zeippen}}{2000}]{Storey00}
{Storey} P.~J.,  {Zeippen} C.~J.,  2000, \mn@doi [\mnras]
  {10.1046/j.1365-8711.2000.03184.x}, \href
  {http://adsabs.harvard.edu/abs/2000MNRAS.312..813S} {312, 813}

\bibitem[\protect\citeauthoryear{{Tsamis}, {Walsh}, {Pequignot}, {Barlow},
  {Liu}  \& {Danziger}}{{Tsamis} et~al.}{2007}]{Tsamis07}
{Tsamis} Y.~G.,  {Walsh} J.~R.,  {Pequignot} D.,  {Barlow} M.~J.,  {Liu} X.-W.,
    {Danziger} I.~J.,  2007, The Messenger, \href
  {http://adsabs.harvard.edu/abs/2007Msngr.127...53T} {127, 53}

\bibitem[\protect\citeauthoryear{{Tylenda}, {Acker}, {Stenholm}  \&
  {Koeppen}}{{Tylenda} et~al.}{1992}]{Tylenda92}
{Tylenda} R.,  {Acker} A.,  {Stenholm} B.,   {Koeppen} J.,  1992, \aaps, \href
  {http://adsabs.harvard.edu/abs/1992A%26AS...95..337T} {95, 337}

\bibitem[\protect\citeauthoryear{{Tylenda}, {Acker}  \& {Stenholm}}{{Tylenda}
  et~al.}{1993}]{Tylenda93}
{Tylenda} R.,  {Acker} A.,   {Stenholm} B.,  1993, \aaps, \href
  {http://adsabs.harvard.edu/abs/1993A%26AS..102..595T} {102, 595}

\bibitem[\protect\citeauthoryear{{V{\'a}zquez}}{{V{\'a}zquez}}{2012}]{Vazquez12}
{V{\'a}zquez} R.,  2012, \mn@doi [\apj] {10.1088/0004-637X/751/2/116}, \href
  {http://adsabs.harvard.edu/abs/2012ApJ...751..116V} {751, 116}

\bibitem[\protect\citeauthoryear{{Walborn} \& {Howarth}}{{Walborn} \&
  {Howarth}}{2000}]{Walborn20}
{Walborn} N.~R.,  {Howarth} I.~D.,  2000, \mn@doi [\pasp] {10.1086/317708},
  \href {http://adsabs.harvard.edu/abs/2000PASP..112.1446W} {112, 1446}

\bibitem[\protect\citeauthoryear{{Weidmann} \& {Gamen}}{{Weidmann} \&
  {Gamen}}{2011}]{Weidmann11}
{Weidmann} W.~A.,  {Gamen} R.,  2011, \mn@doi [\aap]
  {10.1051/0004-6361/200913984}, \href
  {http://adsabs.harvard.edu/abs/2011A%26A...526A...6W} {526, A6}

\bibitem[\protect\citeauthoryear{{Weidmann}, {M{\'e}ndez}  \&
  {Gamen}}{{Weidmann} et~al.}{2015}]{Weidmann15}
{Weidmann} W.~A.,  {M{\'e}ndez} R.~H.,   {Gamen} R.,  2015, \mn@doi [\aap]
  {10.1051/0004-6361/201526096}, \href
  {http://adsabs.harvard.edu/abs/2015A%26A...579A..86W} {579, A86}

\bibitem[\protect\citeauthoryear{{Weinberger}}{{Weinberger}}{1989}]{Weinberger89}
{Weinberger} R.,  1989, \aaps, \href
  {http://adsabs.harvard.edu/abs/1989A%26AS...78..301W} {78, 301}

\bibitem[\protect\citeauthoryear{{Wesson}, {Stock}  \& {Scicluna}}{{Wesson}
  et~al.}{2012}]{Wesson12}
{Wesson} R.,  {Stock} D.~J.,   {Scicluna} P.,  2012, \mn@doi [\mnras]
  {10.1111/j.1365-2966.2012.20863.x}, \href
  {http://adsabs.harvard.edu/abs/2012MNRAS.422.3516W} {422, 3516}

\bibitem[\protect\citeauthoryear{{Zijlstra}, {Pottasch}  \&
  {Bignell}}{{Zijlstra} et~al.}{1989}]{Zijlstra89}
{Zijlstra} A.~A.,  {Pottasch} S.~R.,   {Bignell} C.,  1989, \aaps, \href
  {http://cdsads.u-strasbg.fr/abs/1989A%26AS...79..329Z} {79, 329}

\makeatother
\end{thebibliography}

\end{document}